\documentclass[twoside]{elsart}

\pdfoutput=1

\usepackage[pdftex]{graphicx}
\usepackage{amsmath}
\usepackage{amssymb}
\usepackage{amsfonts}
\usepackage[pdftex]{hyperref}
\usepackage{color}

\newcommand{\impliesn}{$\mathbf{\Rightarrow}$}
\newcommand{\SEPUCfig}[4]{%
\centering
\begin{tabular}{c@{\hspace{1mm}}c@{\hspace{1mm}}c@{\hspace{1mm}}c@{\hspace{1mm}}c}%
Colour image && Binary image && SEPUC \\%
\begin{minipage}[c]{#2\textwidth}%
\includegraphics[width=\textwidth]{#1a}%
\end{minipage}%
& %
\begin{minipage}[c]{.03\textwidth}%
\impliesn%
\end{minipage}%
& %
\begin{minipage}[c]{#3\textwidth}%
\includegraphics[width=\textwidth]{#1b}%
\end{minipage}%
& %
\begin{minipage}[c]{.03\textwidth}%
\impliesn%
\end{minipage}%
& %
\begin{minipage}[c]{#4\textwidth}%
\includegraphics[width=\textwidth]{#1c}%
\end{minipage}%
\end{tabular}%
}

\newcommand\de[1]{\,{\mathrm d}#1}

\newcommand{\bmath}[1]{\mbox{\boldmath$#1$}}
\newcommand{\tenss}[1]{\bmath{#1}}                   

\newcommand{\trn}{{\sf ^T}}

\newcommand{\measure}[1]{|#1|}
\newcommand{\evek}[1]{\{\mathsf{#1}\}}
\newcommand{\emtrx}[1]{\left[\mathsf{#1}\right]}
\newcommand{\strain}{\varepsilon}

\newcommand{\on}{\mbox{ on }}

\newcommand{\abs}[1]{|#1|}

\journal{Materials and Structures}
\pubyear{2007}
\firstpage{1}

\begin{document}                                                                                   

\begin{frontmatter}

\title{Analysis of coupled heat and moisture transfer in masonry structures}
\author[cideas]{Jan S\'{y}kora},
\author[mech]{Jan Vorel},
\author[cideas]{Tom\'{a}\v{s} Krej\v{c}\'{\i}},
\author[mech,cideas]{Michal \v{S}ejnoha}
\author[mech,cideas]{and Ji\v{r}\'{\i} \v{S}ejnoha}
\address[mech]{Department of Mechanics, Faculty of Civil
  Engineering, Czech Technical University in Prague, Th\' akurova 7,
  166 29 Prague 6, Czech Republic} 
\address[cideas]{Centre for Integrated Design of Advances Structures,
  Th\' akurova 7, 166 29 Prague 6, Czech Republic}

\begin{abstract}
Evaluation of effective or macroscopic coefficients of thermal
conductivity under coupled heat and moisture transfer is presented.
The paper first gives a detailed summary on the solution of a simple
steady state heat conduction problem with an emphasis on various types
of boundary conditions applied to the representative volume element --
a periodic unit cell. Since the results essentially suggest no
superiority of any type of boundary conditions, the paper proceeds
with the coupled nonlinear heat and moisture problem subjecting the
selected representative volume element to the prescribed
macroscopically uniform heat flux. This allows for a direct use of the
academic or commercially available codes.  Here, the presented results
are derived with the help of the SIFEL~\footnote{More information
  available at http://mech.fsv.cvut.cz/web/?page=software} (SIimple
Finite Elements) system.
\end{abstract}


\end{frontmatter}

\section{Introduction}\label{sec:intro}
An extensive experimental and numerical analysis of Charles Bridge in
Prague has been performed only recently to identify the most severe
external actions on the bridge, see
e.g.~\cite{Novak:ES:2007,cc2005:sykora,cc2005:sejnoha}. Among others
the loading due to spatially and temporarily varying temperature and
moisture changes appeared to be of paramount importance as these
effects proved as crucial factors responsible for the nucleation and
further development of cracks in the bridge. 
Owing to the complexity of historical masonry structures the problem
is often addressed in a strictly uncoupled format. Distribution of the
temperature field in the macroscopic structure needed for the actual
non-linear thermo-mechanical analysis can be obtained from an
independent solution of the macroscopic steady-state heat conduction
problem~\cite{Novak:ES:2007}. This step, however, requires the
knowledge of the macroscopic coefficients of thermal conductivity, and
possibly their influence on actual temperature and moisture content -
the principal objective of the research presented.

When dealing with these problems, the application of homogenization
techniques is inevitable~\cite{Ozdemir:IJNME:2006}. Solving a set of
problem equations on a meso-scale (a composition of~stone blocks and
mortar) provides us with up-scaled macroscopic equations. They
include a number of effective (macroscopic) transport parameters,
which are necessary for a detailed analysis of the state of a
structure as a whole. A~reliable methodology of the prediction of
these quantities is one of the main goals of~our contribution. Any
(multi-scale) approach to coupled heat and moisture transfer draws on
a cogent description of transport phenomena. An extensive review of
this topic can be found in~\cite{Cerny:2002}. Averaging theories (a
micromechanics-based approach), primarily formulated
in~\cite{Hassanizadech:AWR:1979:1,Hassanizadech:AWR:1979:2}, can be
regarded as a counterpart to phenomenological ones (a
macromechanics-based approach),
see~\cite{DeBoer:AMR:1996,Biot:JAP:1941}. Both approaches are
explained in detail in~\cite{LewisSchrefler:98:2E}. Contrary to
these trends, phenomenological models are still preferred to
averaging ones, namely calculating the heat and moisture transfer in
building materials.

As outlined in~\cite{Cerny:2002}, the models for the description of
water and water vapor transfer can broadly be classified into three
main categories, namely convection models, diffusion models and hybrid
models. The recognized convection model is that of Philip and de
Vries, see e.g.~\cite{Philip:TMGU:1957}. A variety of diffusion models
were developed based on Krischer's original
version~\cite{Krischer:1963}. A certain drawback of this category lies
in the absence of cross-effects between the heat and moisture
transport. Krischer's model was improved by many authors. Let us at
least introduce K\"{u}nzel and Kiessl's
model~\cite{Kunzel:IJHMT:1997}. Because of the lack of space it is
impossible to comment on further approaches such as hybrid models or
complex models that require the application of irreversible
thermo-mechanics,~\cite{Cerny:2002}.

While models for transport processes have been developed during
several decades, the computational methods for multi-scale modeling of
these processes in masonry on meso and macro scales have emerged only
recently. Moreover, most of them are prevailingly confined to the
effective macroscopic description for heat conduction and employ the
perturbation method to describe the fluctuations of temperature
throughout the heterogeneous material. Different situations are
analyzed using the homogenization method, which lead to different
macroscopic descriptions in~\cite{Auriault:IJHMT:1983}.  Boutin's
model analyzing the microstructural influence on heat conduction
belongs to the same category~\cite{Boutin:IJHMT:1995}. It is shown
that the higher order terms introduce successive gradients of
temperature and tensor characteristics of the microstructure, which
result in non-local effects. An original approach to homogenization of
transient heat transfer for some composite materials is proposed
in~\cite{Kaminski:IJES:2003}. In that paper, the stochastic second
moment perturbation method is used in conjunction with the finite
element method. Probably the most complex multi-scale analysis for
pure heat transfer in heterogeneous solids is offered
in~\cite{Ozdemir:IJNME:2006}. The authors established a macro to micro
transition in terms of~the applied boundary conditions and likewise a
micro to macro transition formulated in the form of consistent
averaging relations. See also~\cite{Tomkova:2007} for a similar study
with applications to textile composites.

Homogenization strategies for coupled heat and mass transfer are
rather an exception, see e.g.~\cite{Kaminski:IJES:2003}, and more or
less belong to the modeling of a micro to meso rather than a meso to
macro transition and vice versa. Therein, the condition for a
non-homogenizable situation, i.e. when it is impossible to find a
macroscopic equivalent description, is also addressed. 

While setting up the numerical solution of the homogenization problem
is relatively straightforward, its execution may prove rather
complicated if only standard (academic or commercial) finite element
codes are available. Therefore, apart from the determination of the
influence of temperature and moisture content on effective thermal
conductivities, the present contribution will also be concerned with
a strategy of predicting the desired effective properties when employing
such numerical tools that are unable to directly meet the specifics of the
homogenization problem formulated in the framework of the first-order
homogenization approach. To combine the two goals of this paper, the
following topics will be discussed:
\begin{itemize}
\item
  In Section~\ref{sec:fundamentals} we open the subject by
  reviewing the basic formulas related to first-order homogenization
  clearly identifying the essential similarities between the
  mechanical and heat conduction problem. Therein and also in
  Section~\ref{sec:resistivity} the steady state conditions are
  assumed in the derivation of individual equations. Certain
  restrictions on the local temperature field to comply with the
  assumed macroscopic heat flux or temperature gradient being uniform
  over a certain representative volume element (RVE)
  are summarized.  

\item
  Numerical treatment of the theoretical formulation is then
  presented in Section~\ref{sec:resistivity} to provide estimates of
  the effective thermal conductivities by solving a simple steady
  state heat conduction problem. The main objective of this section is
  to address the effect of boundary conditions on the resulting
  homogenized properties. The fact that these properties are
  essentially invariant with respect to the choice of particular
  boundary conditions opens the way to the solution of the coupled
  heat and moisture problem using the SIFEL finite element
  code, a typical representative of the finite element software not
  specifically developed for the solution of the homogenization
  problem. Implementation of the loading and boundary conditions into
  such codes is therefore also discussed.

\item
  The essentials of theoretical formulation as well as some
  numerical results are presented in Section~\ref{sec:moisture}
  clearly illustrating the effect of moisture on thermal conductivity
  and their dependence on varying material parameters, such as
  relative humidity and initial temperature. Here, unlike the previous
  sections, the analysis is carried out under transient
  conditions. The presented results suggest that the predicted
  effective thermal conductivities are almost insensitive to the
  changes in the macroscopic temperature gradient. This is rather
  appealing, particularly in view of the expected use of the
  homogenized properties in the solution of an independent macroscopic
  steady state heat conduction problem to provide space variation of
  the temperature field within the macroscopic structure (e.g. Charles
  Bridge in Prague), which in turn is used in the actual mechanical
  analysis~\cite{Novak:ES:2007}.

\item
  Finally, the essential findings are summarized in
  Section~\ref{sec:conclusions}. 
\end{itemize}

In Section~\ref{sec:fundamentals}, the standard indicial notation is
used such that $t_{,i}$ represents a derivative of a scalar quantity
$t$ with respect to a coordinate $x_i$. The summation over the repeated
indexes is assumed. As more typical of the discretized form of the
governing equations using the finite element method, we adopt in
Section~\ref{sec:resistivity} the matrix-vector notation such that the
symbol $\evek{a}$ is reserved for a vector and $\emtrx{A}$ is employed
for a matrix representation, respectively
~\cite{Bittnar:1996:NMM}. The symbol $\evek{\nabla{\it t}}=
\displaystyle{\left\{\frac{\partial{t}}{\partial{x}},\frac{\partial{t}}{\partial{y}},\frac{\partial{t}}{\partial{z}}\right\}^{\sf
    T}}$ then represents the gradient of $t$ and $\evek{\cdot}^{\sf
  T},\emtrx{\cdot}^{\sf T}$ stands for the transpose vector or matrix,
respectively. Finally, the symbol $\dot{t}$ used in
Section~\ref{sec:moisture} represents the time derivative of a given
quantity $t$.

\begin{figure} [ht]
\begin{center}
\begin{tabular}{c}
\includegraphics*[width=70mm,keepaspectratio]{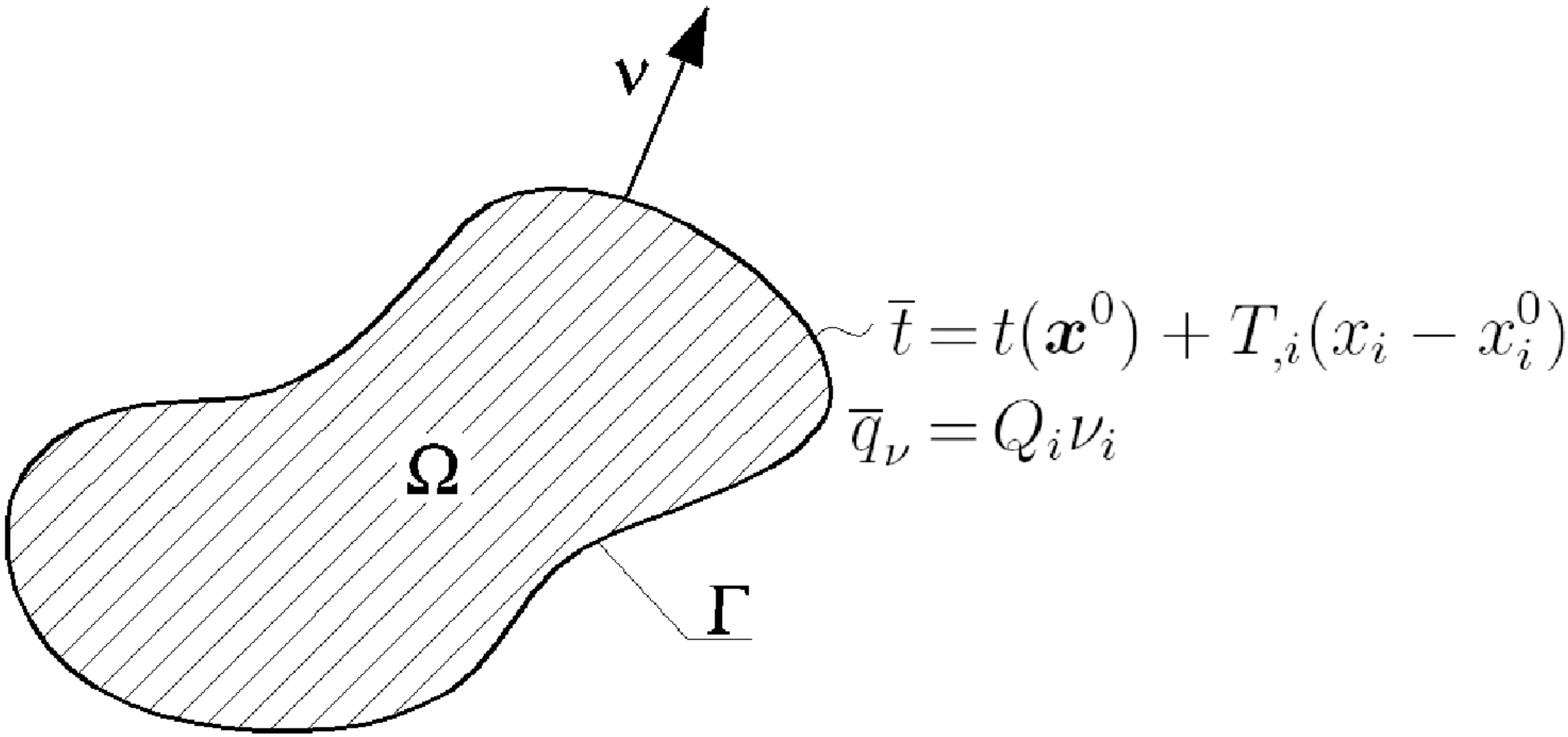}
\end{tabular}
\caption{A representative volume element with macroscopic loading conditions}
\label{fig:RVE}
\end{center}
\end{figure}

\section{Fundamentals of homogenization}\label{sec:fundamentals}
Consider a representative volume element $\measure{\Omega}$ in
Fig.~\ref{fig:RVE} loaded over the boundary $\Gamma$ either by the
prescribed temperature $\overline{t}$ derived from a uniform macroscopic
temperature gradient $T_{,i}$ or by a normal heat flux
$\overline{q}_{\nu}$ consistent with the macroscopic uniform heat flux
$Q_i$ such that
\begin{eqnarray}
\overline{t} &=& t(\tenss{x}^0) + T_{,i}(x_i-x_i^0)\hspace{1cm}\on\Gamma,\nonumber \\ 
\overline{q}_{\nu} &=& Q_i\nu_i\hspace{3.72cm}\on\Gamma,\label{eq:bcmacro} 
\end{eqnarray}
where $\nu_i$ represents the $i$-th component of the outward unit
normal.~\footnote{Note that in the case of nonlinear analysis the total
quantities would have to be substituted by their respective
increments.}

In such a case, the local temperature field admits the following
decomposition 
\begin{equation}
t(\tenss{x}) = t(\tenss{x}^0) + T_{,i}(x_i - x_i^0) +t^*(\tenss{x}),
\label{eq:t}
\end{equation}
where $T_{,i}$ are components of the macroscopically uniform
temperature gradient vector and $t^*(\tenss{x})$ is the fluctuation of
the local temperature field.
Finally, the temperature $t(\tenss{x}^0)$ at the reference point
$\tenss{x}^0$ is introduced to specify the distribution of the local
temperature field uniquely. This term plays an important role when a
fully coupled multi-scale analysis is considered and/or when the
instantaneous effective thermal conductivities depend on the current
temperature. If, however, we seek only for temperature independent
homogenized properties, like in Section~\ref{sec:resistivity}, this
term becomes irrelevant and can be safely omitted.  

Eq.~\eqref{eq:t} immediately follows from the relation between the
temperature gradients of individual fields
\begin{equation}
t_{,i}(\tenss{x}) = T_{,i} + t^*_{,i}(\tenss{x})\label{eq:local-grad-temp}.
\end{equation}
The micro-temperature gradient averaged over the volume $\measure{\Omega}$
of the representative volume element
\begin{equation}
\langle t_{,i} \rangle = \frac{1}{\measure{\Omega}}\int_{\Omega} t_{,i}(\tenss{x}) \de\Omega(\tenss{x}) = T_{,i} +
\frac{1}{\measure{\Omega}}\int_{\Omega} t^*_{,i}(\tenss{x})
\de\Omega(\tenss{x}),\label{eq:avggT}
\end{equation}
yields a scale transition relation, see
e.g.~\cite{Ozdemir:IJNME:2006},
\begin{equation}
\langle t^*_{,i} \rangle = \frac{1}{\measure{\Omega}}\int_{\Omega}
t^*_{,i}(\tenss{x}) \de\Omega(\tenss{x}) =
\frac{1}{\measure{\Omega}}\int_{\Gamma} t^*(\tenss{x})\nu_i(\tenss{x})
\de\Gamma(\tenss{x}) = 0. \label{eq:t*}
\end{equation}
The boundary integral disappears providing either the fluctuation part
of the temperature field equals zero (in the case of fully prescribed
temperature and/or normal heat flux boundary conditions) or the
periodic boundary conditions, i.e. the same values of $t^*$ on
opposite sides of a rectangular RVE, are enforced on $\Gamma$.

When discussing the micro to macro scale transition it is worth
mentioning an analogy between the basic quantities related to the
heat conduction problems (i.e. the negative values of
temperature gradients $(-t_{,i})$ or $(-T_{,i})$ and fluxes $q_i$
or $Q_i$ as their conjugate measures) and the corresponding
quantities applied to mechanical problems (strains
$\varepsilon_{ij}$ or $E_{ij}$ and the conjugate stress measures
$\sigma_{ij}$ or $\Sigma_{ij}$). While homogenization of
mechanical problems benefits from the Hill lemma
\begin{equation}
\langle {\varepsilon}_{ij} \sigma_{ij} \rangle = {E}_{ij}
\Sigma_{ij}, \label{eq:Hill}
\end{equation}
where $E_{ij}, \Sigma_{ij}$ are macroscopically uniform strain and
stress tensors and $\strain_{ij}, \sigma_{ij}$ are their local
counterparts, 
the Fourier inequality~\cite{Malvern:1969:IMCM} is used
to obtain an equivalent representation of Eq.~\eqref{eq:Hill} given by
%
\begin{equation}
\langle (-t_{,i}) q_i \rangle = -T_{,i} Q_i, \label{eq:tq}
\end{equation}
which is applicable in the case of homogenization of the heat
conduction problems.

Following the transformation of the volume integral in
Eq.~\eqref{eq:t*} to its boundary equivalent it is useful to convert
the left-hand side of Eq.~\eqref{eq:tq} in a similar way
\begin{eqnarray}
\langle t_{,i}q_i \rangle & = & \frac{1}{\measure{\Omega}} \int_{\Omega}
t_{,i}(\tenss{x}) q_i (\tenss{x}) \de \Omega (\tenss{x}) =
\frac{1}{\measure{\Omega}} \int_{\Gamma} t(\tenss{x}) q_i (\tenss{x}) \nu_i (\tenss{x}) \de \Gamma (\tenss{x}) = \nonumber \\
 & = & \frac{1}{\measure{\Omega}} \int_{\Gamma} t(\tenss{x}) q_{\nu} (\tenss{x}) \de \Gamma (\tenss{x}).
\label{eq:tq2}
\end{eqnarray}
The term
\begin{equation}
- \frac{1}{\Omega} \int_{\Omega} t(\tenss{x}) q_{i,i} (\tenss{x})
\de \Omega (\tenss{x}),
\end{equation}
has been omitted in Eq.~\eqref{eq:tq2} because of the balance of heat
(i.e. $q_{i,i} = 0$ under steady state conditions and the absence of
internal heat sources).

The same boundary conditions, which satisfy Eq.~\eqref{eq:t*}, lead to
the equivalence of the volume averaged microscopic heat flux $q_i$ and
the macroscopic heat flux $Q_i$. To prove this, the well-known
identity
\footnote{$ \int_{\Gamma} x_i q_{\nu} (\tenss{x}) \de \Gamma (\tenss{x}) = \int_{\Gamma} x_i q_j (\tenss{x}) \nu_j (\tenss{x}) \de \Gamma (\tenss{x}) = \int_{\Omega}
(x_i q_j (\tenss{x}))_{,j} \de \Omega (\tenss{x}) =
\int_{\Omega} ( \delta_{ij} q_j (\tenss{x}) + x_i \underbrace{q_j (\tenss{x})_{,j}}_{0} ) \de
\Omega (\tenss{x}) = \int_{\Omega} q_i (\tenss{x}) \de \Omega
(\tenss{x})$}
\begin{equation}
\int_{\Omega} q_i (\tenss{x}) \de \Omega (\tenss{x}) =
\int_{\Gamma} x_i q_{\nu} (\tenss{x}) \de \Gamma (\tenss{x}),
\label{eq:intqi}
\end{equation}
is utilized. The following three situations can be distinguished:
\begin{itemize}
\item In the case of fully prescribed temperature boundary conditions,
the subsequent combination of Eqs.~\eqref{eq:t},~\eqref{eq:tq2}
and~\eqref{eq:intqi}, in conjunction with the balance condition
\begin{equation}
\int_{\Gamma} q_{\nu} (\tenss{x}) \de \Gamma (\tenss{x}) = 0,
\end{equation}
yields
\begin{eqnarray}
\langle t_{,i} q_i \rangle & = & \frac{t(\tenss{x}^0) }{\measure{\Omega}}
\int_{\Gamma} q_{\nu} (\tenss{x}) \de \Gamma (\tenss{x}) +
\frac{T_{,i}}{\measure{\Omega}} \int_{\Gamma} ( x_i - x^0_i ) q_{\nu} (\tenss{x}) \de \Gamma (\tenss{x}) = \nonumber \\
 & = & \frac{T_{,i}}{\measure{\Omega}} \int_{\Omega} q_i (\tenss{x}) \de \Omega (\tenss{x}) = T_{,i} Q_i.
\label{eq:combination}
\end{eqnarray}
This identity implies
\begin{equation}
Q_i = \frac{1}{\measure{\Omega}} \int_{\Omega} q_i (\tenss{x}) \de
\Omega (\tenss{x}). \label{eq:Qi}
\end{equation}
\item A similar situation is met if normal heat flux boundary
conditions are prescribed. Substituting the prescribed macroscopic
flux $Q_i$ into Eq.~\eqref{eq:combination} for $q_i$ immediately
proves the validity of Eq.~\eqref{eq:Qi}.
\item The periodic temperature boundary conditions are typical of
rectangular PUCs. From the transition relation~\eqref{eq:t*}, the
fluctuation temperature $t^*$ must be the same on the opposite
boundaries of PUC. As the anti-periodic normal heat flux applies to
the periodic boundary conditions, Eq.~\eqref{eq:combination} yields,
after certain modifications, the expected result~\eqref{eq:Qi}.
\end{itemize}

\section{Macroscopic conductivity and resistivity matrices}\label{sec:resistivity}
Before proceeding with the analysis of a complex coupled nonlinear
heat and moisture conduction problem we review the basic steps for the
evaluation of macroscopic conductivity and resistivity matrices
through the solution of a simple steady state heat conduction problem
using the finite element method. This allows us to examine the
influence of various boundary conditions, discussed in the previous
section, on the predicted homogenized properties and further exploit
these results in the next section when solving the coupled problem.

As intimated in the previous section, compare
Eqs.~\eqref{eq:Hill}~-~\eqref{eq:tq}, the macroscopic conductivity and
resistivity matrices may be derived by analogy to the homogenized
stiffness and compliance matrices which apply to mechanical
analyses. Therefore, we first discuss the numerical treatment of
Eq.~\eqref{eq:Hill}, which allows the RVE being directly loaded by
prescribed macroscopic uniform strain or stress fields $E_{ij}$ or
$\Sigma_{ij}$, respectively. The analogous loading conditions
(macroscopic temperature gradient $T_{,i}$ or macroscopic heat flux
$Q_i$) that apply to heat conduction problem are examined next. Such
loading conditions are, however, difficult to introduce directly into
most finite element codes including the SIFEL program. Instead, the
loading conditions of the type~\eqref{eq:bcmacro} are needed. This, on
the other hand, calls for a special treatment of the boundary
conditions associated with a fluctuation part of the local temperature
field $t^*$. To elucidate this subject, a brief comment is given in
the last part of this section.

\subsection{Searching the solution in terms of the fluctuation part of the local temperature}
To begin, consider a rectangular RVE (particular definition of an RVE
will be given later in Section~\ref{subsec:simpleexample}) subjected
either to macroscopic uniform strain $\evek{E}$ or stress
$\evek{\Sigma}$. Henceforth, as already mentioned in the introductory
part, the standard vector-matrix notation is adopted. Limiting our
attention to linear elastic theory, thus allowing for small
displacements, rotations and strains only, provides the virtual work
representation of Eq.~\eqref{eq:Hill} in the form
\begin{equation}
\langle\evek{\delta\varepsilon}\trn\evek{\sigma}\rangle = \evek{\delta{E}}\trn\evek{\Sigma}.\label{eq:Hill-2}
\end{equation}
where the discretized form of the local strain, in analogy with
Eq.~\eqref{eq:local-grad-temp}, and stress fields are provided by
\begin{equation}
\evek{\strain} = \evek{E} + \emtrx{B_u}\evek{r^*_u},\quad\evek{\sigma}=\emtrx{L}\evek{\strain}.\label{eq:local-strain}
\end{equation}
In Eq.~\eqref{eq:local-strain} $\evek{r_u^*}$ is the vector of the
nodal values of the fluctuation displacement field being periodic (the
same values of $\evek{r_u^*}$ on opposite sides of an RVE),
$\emtrx{B_u}$ stores derivatives of the displacement shape functions
and $\emtrx{L}$ is the microscopic material stiffness matrix. This
equation readily suggests that the solution of Eq.~\eqref{eq:Hill-2}
is searched in terms of fluctuation rather than actual
displacements. Next, introducing Eq.~\eqref{eq:local-strain} into
Eq.~\eqref{eq:Hill-2} gives
\begin{equation}
\left<\left(\evek{\delta{E}}\trn+\evek{\delta{r_u^*}}\trn\emtrx{B_u}\trn\right)\emtrx{L}\left(\evek{E}+\emtrx{B_u}\evek{r_u^*}\right)\right>
= \evek{\delta{E}}\trn\evek{\Sigma},\label{eq:Hill-3}
\end{equation}
to be satisfied for all kinematically admissible strains
$\evek{\delta{E}}$ and displacements $\evek{\delta{r_u^*}}$.

Let $\evek{E}$ be now the prescribed macroscopic strain vector. The
above equation then reduces to (note that $\evek{\delta{E}}=\evek{0}$)
\begin{equation}
 \langle\emtrx{B_u}\trn\emtrx{L}\emtrx{B_u}\rangle
\evek{r_u^*} = -\langle \evek{B_u}\trn\emtrx{L}\rangle\evek{E}, \label{eq:powers}
\end{equation}
to be solved for unknown nodal values of the fluctuation part of the
displacement field $\evek{r_u^*}$. The macroscopic constitutive law
can be then expressed as
\begin{equation}
\evek{\Sigma} = \langle\evek{\sigma}\rangle = \langle\emtrx{L}(\evek{E} + \evek{B_u}\evek{r_u^*})
\rangle = \emtrx{L}^{hom}\evek{E},\label{eq:hom-L}
\end{equation}
%
where $\emtrx{L}^{hom}$ is the homogenized (macroscopic) material
stiffness matrix. Assuming e.g. plane-stress conditions the
coefficients of the $3\times{3}$ matrix $\emtrx{L}^{hom}$ are defined
as volume averages of the local fields derived from the solution of
three successive elasticity problems. To that end, the representative
volume element is loaded, in turn, by each of the three components of
$\evek{E}$, while the other two vanish. The volume stress averages,
Eq.~\eqref{eq:hom-L}, normalized with respect to $\evek{E}$ then
furnish individual columns of $\emtrx{L}^{hom}$. Note that
introduction of the periodic boundary conditions is, in this
particular case, relatively simple as it is only required to assign,
in the FEM terminology, the same code numbers to corresponding degrees
of freedom of $\evek{r_u^*}$ on opposite sides of the RVE. To avoid
the rigid body motion and also as a result of the application of
periodic boundary conditions the rectangular RVE is fixed at all
corners.
 
If, on the other hand, the macroscopic stress $\evek{\Sigma}$ is
prescribed (note that in this case $\evek{\delta{E}}\ne\evek{0}$) a
similar procedure is employed to arrive at a set of the following
equations for unknown ${\evek{E}}$ and ${\evek{r}_u^*}$
\begin{eqnarray}
\langle\emtrx{L}\rangle\evek{E} + \langle\emtrx{L}\emtrx{B_u}\rangle\evek{r_u^*} & = & \evek{\Sigma}, \nonumber \\
\langle\emtrx{L}\emtrx{B_u}\rangle\trn\evek{E} + \langle\emtrx{B_u}\trn\emtrx{L}\emtrx{B_u}\rangle
\evek{r_u^*} & = & \evek{0}.\label{eq:Hill-4}
\end{eqnarray}
Here, solving for $\evek{E}$ from the three independent elasticity
problems and again normalizing by $\evek{\Sigma}$ readily provides the
macroscopic material compliance matrix $\emtrx{M}^{hom} = (
\emtrx{L}^{hom})^{-1}$ as
\begin{equation}
\evek{E} = \emtrx{M}^{hom} \evek{\Sigma}.\label{eq:macro}
\end{equation}

Derivation of the effective thermal conductivities and resistivities
will be now presented on the same footing taking advantage of the
analogy between mechanical and heat conduction problem. To do so, we
first write Eq.~\eqref{eq:tq} in the form similar to
Eq.~\eqref{eq:Hill-2}
\begin{equation}
\langle\evek{\delta\nabla{\it t}}\trn\evek{\it q}\rangle = \evek{\delta\nabla{\it T}}\trn\evek{\it Q}.\label{eq:tq-2}
\end{equation}
The discretized form of the local temperature gradient and heat
flux now reads
\begin{equation}
\evek{\nabla{\it t}} = \evek{\nabla{\it T}} + \emtrx{B_t}\evek{r_t^*}, \quad \evek{\it q} = -\emtrx{\chi}\evek{\nabla{\it t}},
\label{eq:nablat}
\end{equation}
where $\evek{r_t^*}$ is now the vector of nodal values of the
fluctuation temperature field, $\emtrx{\chi}$ is the microscopic
conductivity matrix and entries of the matrix $\emtrx{B_t}$ represent
the derivatives of the temperature shape functions.

For the prescribed macroscopic temperature gradient $\evek{\nabla{\it
    T}}$ the equality~\eqref{eq:tq-2} results in (compare with
Eq.~\eqref{eq:powers})
\begin{equation}
\langle\emtrx{B_t}\trn\emtrx{\chi}\emtrx{B_t} \rangle\evek{r_t^*} =
-\langle \emtrx{B_t}\trn\emtrx{\chi}\rangle\evek{\nabla{\it
    T}},\label{eq:tfluc}
\end{equation}
and, successively, in the macroscopic relation for the heat flux
\begin{equation}
\evek{\it Q} = \langle\evek{\it q}\rangle = - \langle\emtrx{\chi}(\evek{\nabla{\it T}} + \emtrx{B_t}\evek{r_t^*})
\rangle = - \emtrx{\chi}^{hom} \evek{\nabla{\it T}},\label{eq:Chihom}
\end{equation}
%
where the homogenized conductivity matrix $\emtrx{\chi}^{hom}$ now
follows, with the expected restrictions to two-dimensional problem,
from the solution of two successive heat conduction problems in
exactly the same way as already outlined for the homogenized stiffness
matrix $\emtrx{L}^{hom}$. The periodic boundary conditions now require
the same fluctuation temperatures to be ensured on opposite sides of
the RVE, see also Fig.~\ref{fig:pbc} and
Eq.~\eqref{eq:periodic-sifel}. In analogy to the mechanical problem a
zero value of $t^*$ is prescribed at all corners of the RVE.  

The prescribed macroscopic heat flux $\evek{\it Q}$ then provides, in
analogy with Eq.~\eqref{eq:Hill-4},
\begin{eqnarray}
-\langle\emtrx{\chi}\rangle \evek{\nabla{\it T}} - \langle\emtrx{\chi}\emtrx{B_t}\rangle\evek{r_t^*} & = & \evek{\it Q}, \nonumber \\
\langle \emtrx{\chi}\emtrx{B_t}\rangle\trn\evek{\nabla{\it T}} + \langle \emtrx{B_t}\trn\emtrx{\chi} \emtrx{B_t} \rangle
\evek{r_t^*} & = & \evek{0},\\
\evek{\nabla{\it T}} &=& - \emtrx{\Psi}^{hom}\evek{\it Q},\label{eq:hom-psi}
\end{eqnarray}
where $\emtrx{\Psi}^{hom} = (\emtrx{\chi}^{hom})^{-1}$ is the
effective (macroscopic) resistivity matrix.

\subsection{Influence of boundary conditions on effective conductivities}\label{subsec:simpleexample}
\begin{figure} [ht!]
\begin{center}
\begin{tabular}{c}
\includegraphics[width=80mm,keepaspectratio]{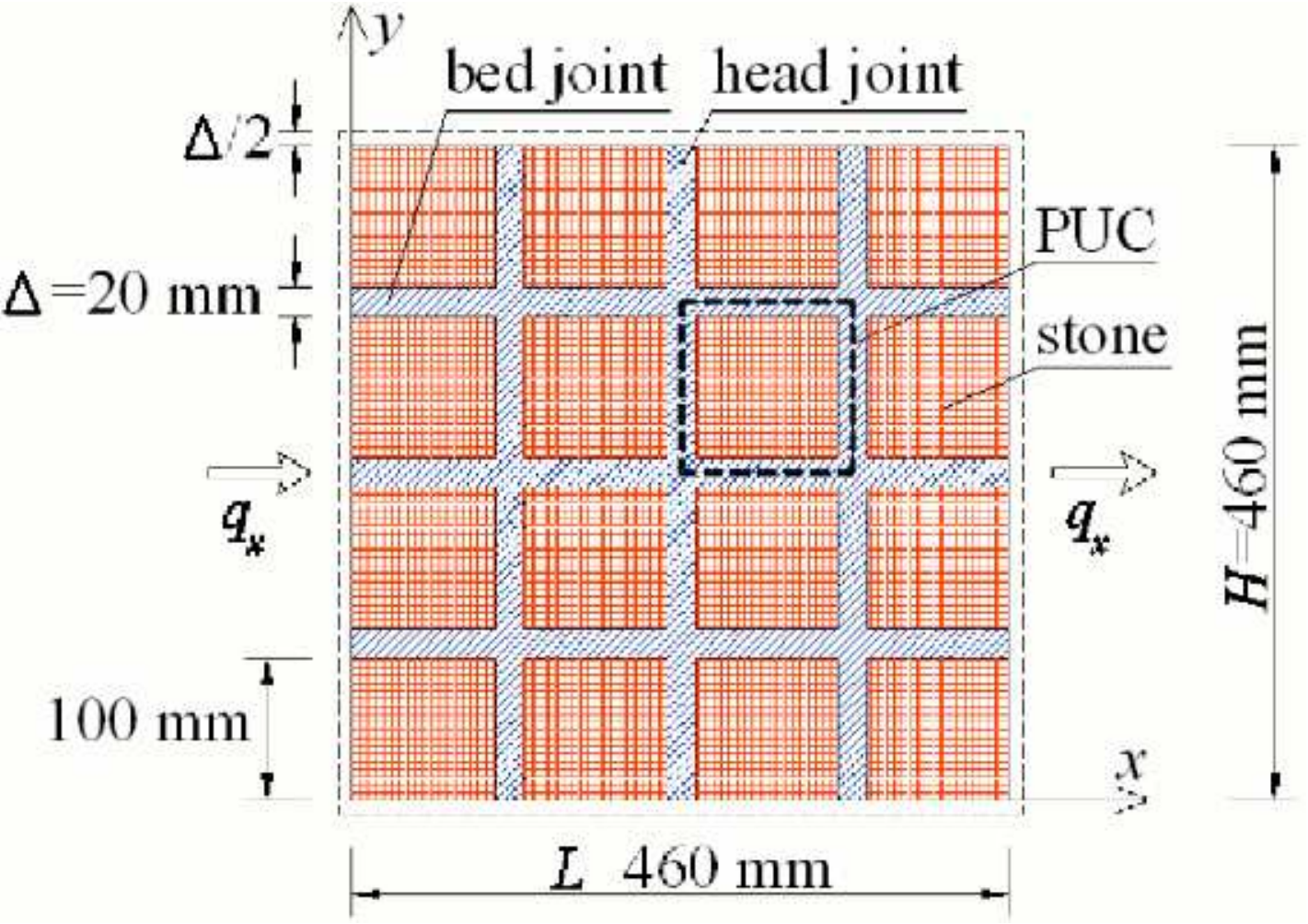}
\end{tabular}
\caption{Meso-structure consisting of stone blocks and mortar joints} 
\label{fig:mesostrucutre}
\end{center}
\end{figure}

To capture the influence of the selected type of boundary conditions
consider an RVE displayed in Fig.~\ref{fig:mesostrucutre}. Such a
meso-structure does not represent a typical masonry bonding, as in
this scheme both the stone blocks and bed and head joints are
regularly and uniformly distributed. It is also worth noting that such
an arrangement is not perfectly periodic due to the disturbance of
periodicity along the boundary. This large RVE has been deliberately
chosen to demonstrate certain edge effects, which may be taken into
account when homogenization is carried out using commercial and/or
academic computer codes. If the RVE was perfectly periodic (i.e. one
half of the boundary mortar joint was added as sketched by the dashed
line in Fig.~\ref{fig:mesostrucutre}), then, due to the infinite
periodicity, any stone block along with the adjacent part of joints
shaded in Fig.~\ref{fig:mesostrucutre} would be sufficient for
calculating the effective value of thermal conductivity.

\begin{table}[ht]
\centering
\begin{tabular}{|c|c|}\hline
mortar & sandstone \\
\hline
0.87	    &1.9\\
\hline
\end{tabular}
\caption{Phase conductivities ${\chi}[\mathrm{Wm^{-1}K^{-1}}]$ at $t=25^0C, \varphi=0\%$}
\label{Tab1}
\end{table}

\begin{figure}
\begin{center}
\begin{tabular}{c}
\includegraphics[width=120mm,keepaspectratio]{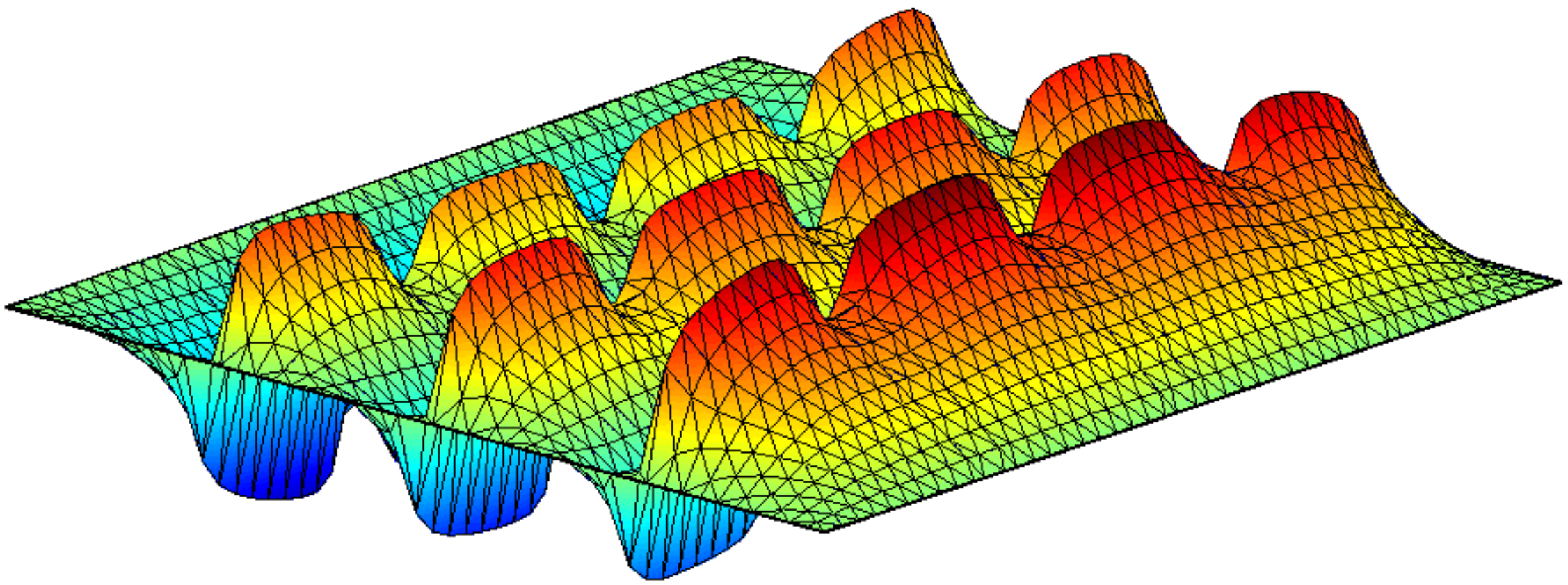}\\
(a)
\end{tabular}
\begin{tabular}{cc}
\includegraphics[width=70mm,keepaspectratio]{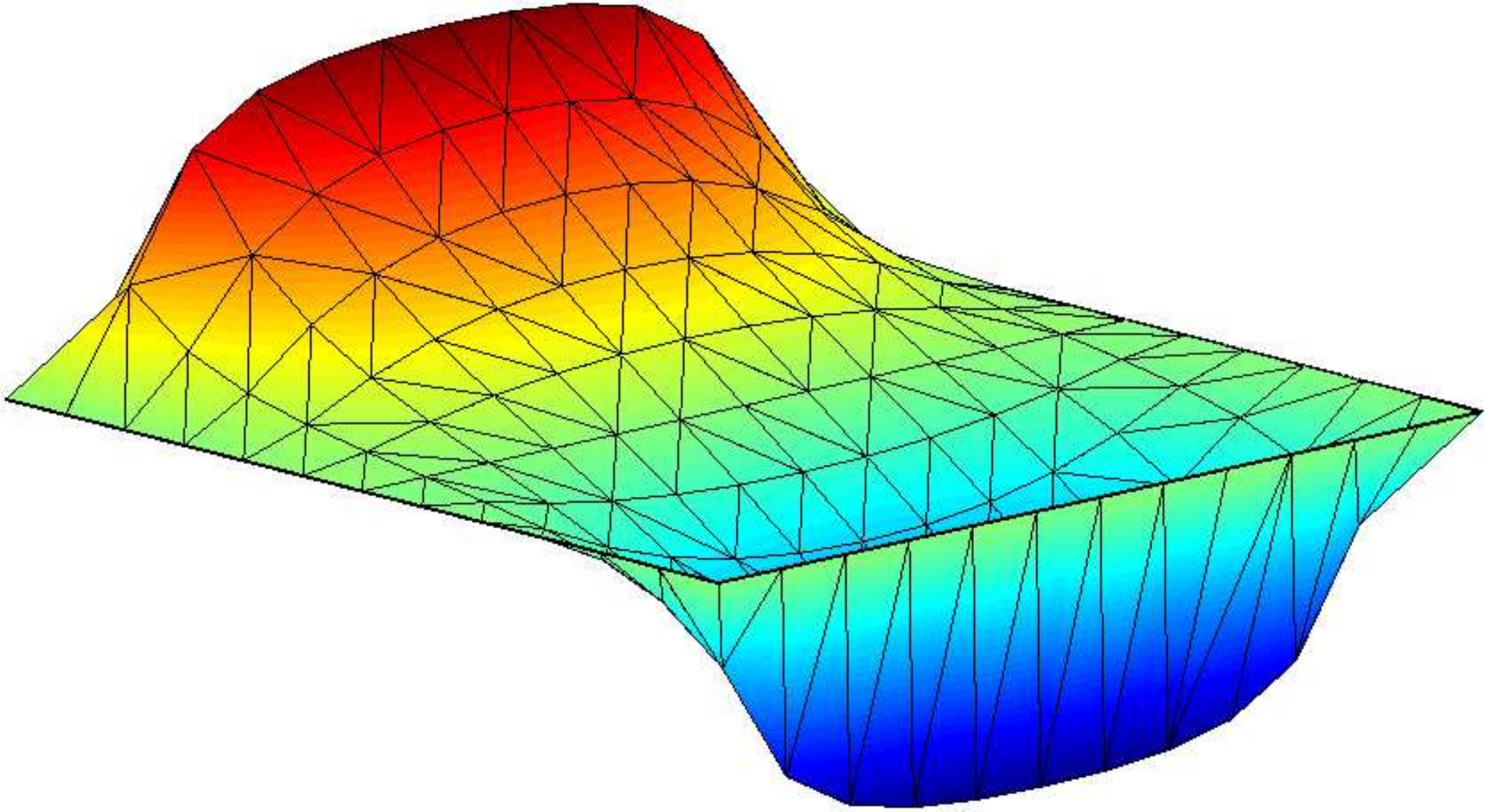}&
\includegraphics[width=70mm,keepaspectratio]{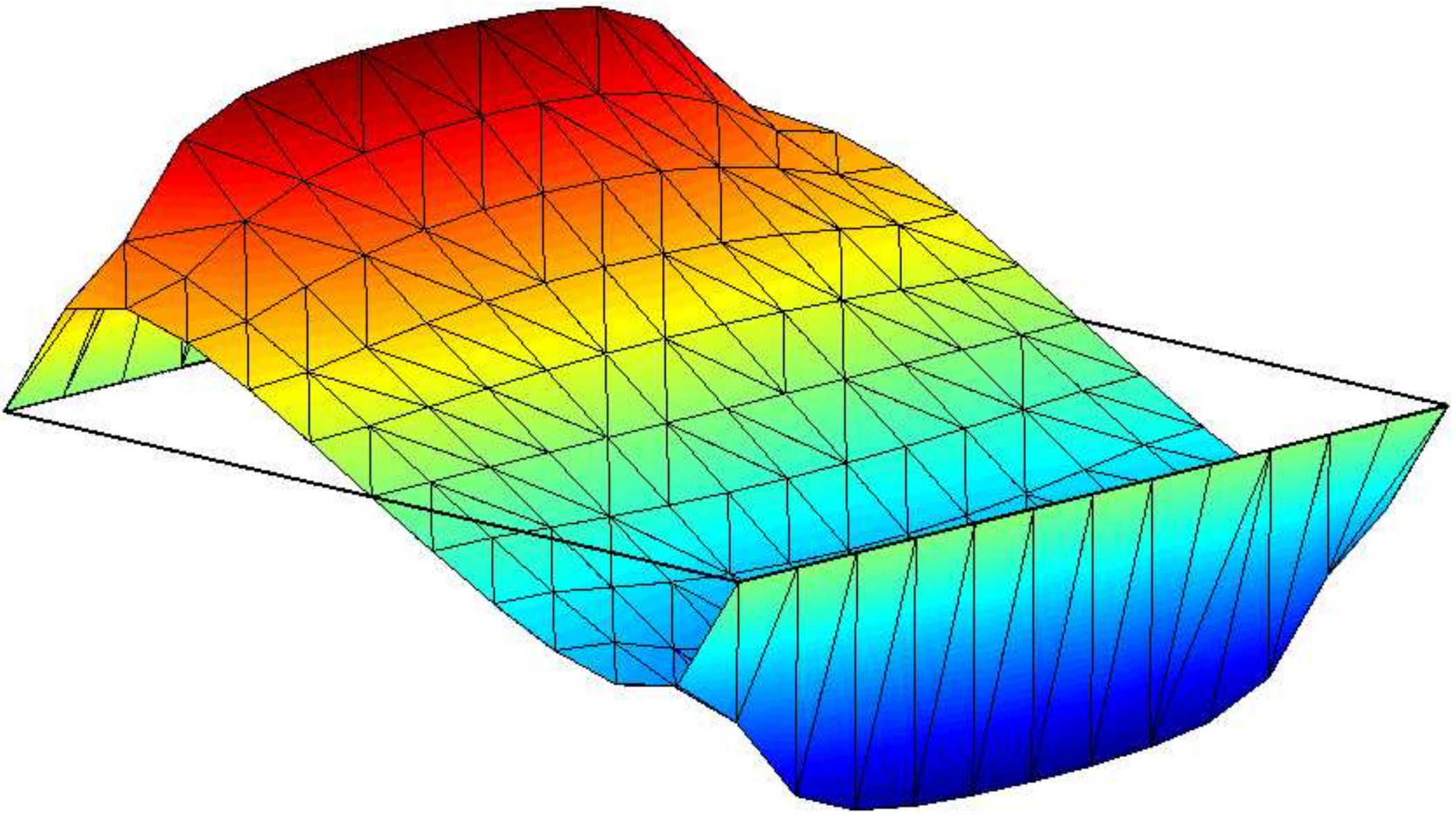}\\
(b)&(c)
\end{tabular}
\caption{Fluctuation temperature field: (a) large RVE and $t^*=0$ (b) PUC and $t^*=0$ (c) PUC and $t^*$-periodic} 
\label{fig:fluctemp}
\end{center}
\end{figure}

In this simple case the coupling effects between the heat and moisture
were omitted. The phase thermal conductivities are listed in
Table~\ref{Tab1}. First, the effect of imperfect periodicity was
studied by loading the large RVE in Fig.~\ref{fig:mesostrucutre}
consisting of 16 stone blocks and surrounding bed and head joints
by a uniform macroscopic temperature gradient $\nabla_x T=1$ while
setting $\nabla_y T=0$. Fully prescribed boundary temperatures were
considered so that $t^{*}=0$. The fluctuation temperature field
displayed in Fig.~\ref{fig:fluctemp}(a) was found from the solution of
Eq.~\eqref{eq:tfluc}. Note that the imperfect periodicity of the RVE
(the edge effect) manifests itself in slightly affecting the
periodicity of the fluctuation field
$t^{*}$. Fig.~\ref{fig:fluctemp}(b) shows similar results derived by
assuming the RVE in terms of the PUC. Finally, the results plotted in
Fig.~\ref{fig:fluctemp}(c) were generated again through the solution
of the unit cell problem but incorporating the periodic boundary
conditions. 
To avoid possible confusion the large RVE, now modified to comply with
perfect periodicity, was finally examined. Both the influence of fully
prescribed boundary temperatures ($t^*=0$ on $\Gamma$) and the
periodic boundary conditions were again studied.

\begin{table}[ht!]
\centering
\begin{tabular}{|c|c|c|c|c|}\hline
Large RVE & Large RVE & Large RVE & PUC & PUC \\
non-periodic & periodic  & periodic & & \\
$t^*=0$ on $\Gamma$ & $t^*=0$ on $\Gamma$ & $t^*$-periodic & $t^*=0$ on $\Gamma$ & $t^*$-periodic\\
\hline
1.563 & 1.48 & 1.47 & 1.48 & 1.47\\
\hline
Voight & 1.58 & \multicolumn{3}{c|}{$\chi_x^{hom} = c^m\chi_x^m+c^s\chi_x^s$}\\
\hline
Reuss & 1.39 &  \multicolumn{3}{c|}{$1/\chi_x^{hom} = c^m/\chi_x^m+c^s/\chi_x^s$}\\
\hline
\end{tabular}
\caption{Effective conductivities ${\chi}^{hom}_x[\mathrm{Wm^{-1}K^{-1}}]$}
\label{Tab2}
\end{table}

The corresponding effective thermal conductivities were obtained
directly from the solution of Eq.~\eqref{eq:tfluc} together with
Eq.~\eqref{eq:Chihom}. Individual values are stored in
Table~\ref{Tab2} clearly suggesting, at least in this particular
example, the invariance of the predicted effective thermal
conductivities on the choice of specific boundary conditions. This
appealing result is further exploited in
Section~\ref{subsec:comcodes}.  
The Voight bound derived from a simple rule of mixture and the Reuss
bound provided by an inverse rule of mixture are presented for
illustration. The volume fractions of mortar $c^m=0.31$ and stones
$c^s=0.69$ correspond to periodic structure 
(PUC in Fig.~\ref{fig:mesostrucutre}).

\subsection{Searching the solution in terms of the local temperature - periodic boundary conditions in commercial codes}\label{subsec:effectiveconductivities}
Recall that the effective thermal conductivities have been obtained
assuming the solution to be derived in terms of the fluctuation part
$t^*$ of the local temperature $t$ when loading an RVE directly by the
prescribed macroscopic temperature gradient $\evek{\nabla{\it T}}$ or
by the macroscopic uniform heat flux $\evek{\it Q}$. The designer,
however, must often rely on the use of standard finite element codes,
either academic or commercial, where the loading is represented in
terms of the prescribed boundary temperatures or fluxes of the
type~\eqref{eq:bcmacro} and the solution is searched directly in terms
of the local temperatures $t(\tenss{x}), \tenss{x}\in\Omega$.

\begin{figure} [ht!]
\begin{center}
\begin{tabular}{c}
\includegraphics[width=70mm,keepaspectratio]{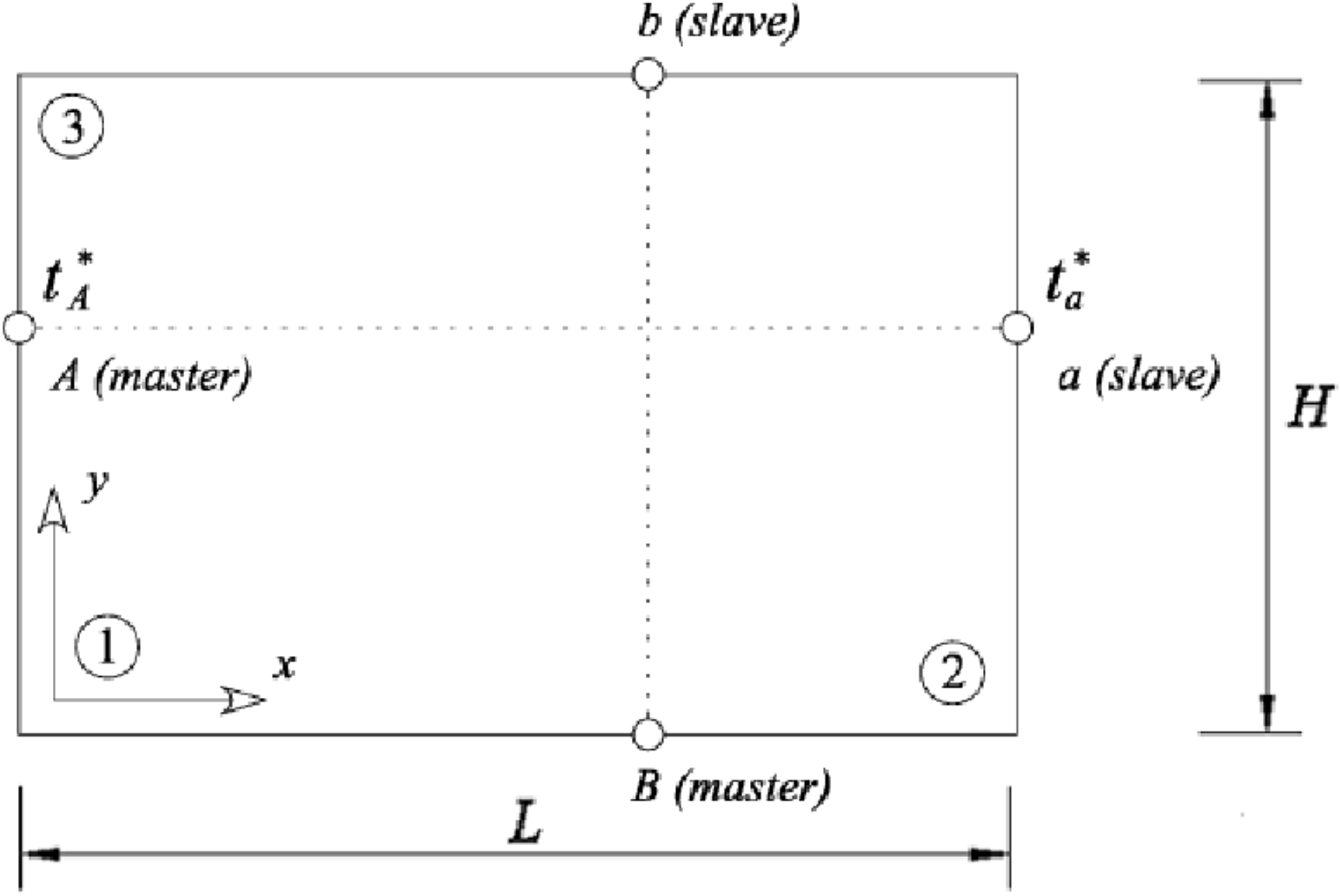}
\end{tabular}
\caption{Conditions of periodicity} \label{fig:pbc}
\end{center}
\end{figure}

Consider now the most simple case illustrated in
Fig.~\ref{fig:mesostrucutre}, when the RVE is loaded by the uniform
heat flux $\abs{q_{\nu}}=q_x=Q_x$ along the two vertical boundaries at $x=0,
x=L$ while no flow boundary conditions are specified
$\abs{q_{\nu}}=q_y=Q_y=0$ along the two horizontal boundaries at $y=0, y=H$.
Eq.~\eqref{eq:Qi} is then immediately satisfied and the horizontal
component of the effective heat conductivity is, in view of
Eqs.~\eqref{eq:hom-psi} and~\eqref{eq:local-grad-temp}, provided by
\begin{equation}
\chi_x^{hom} = -\frac{q_x}{\langle{t}_{,x}\rangle}.\label{eq:chi-hom-gx}
\end{equation}
If no action is taken this result corresponds to the assumption of
$t^*=0$ on $\Gamma$. If, on the other hand, the periodic boundary
conditions are on demand it is necessary proceed as follows. 
Consider again a two-dimensional rectangular RVE with dimensions $H$
and $L$ (see Fig. \ref{fig:pbc}).

Observe that for a pair of points (e.g. $A$ - master and $a$ - slave)
located on the opposite sides of the PUC the following relations hold:
\begin{equation}
t_A = \left( \frac{\partial T}{\partial y} \right) y_A + t^*_A +
t(\tenss{x}^0), \quad t_a = \left( \frac{\partial T}{\partial x}
\right) L + \left( \frac{\partial T}{\partial y} \right) y_a +
t^*_a + t(\tenss{x}^0).\label{eq:local_t-sifel}
\end{equation}
Taking into account the fact that the fluctuation field $t^*$
satisfies the periodicity condition
\begin{equation}
t^*_a = t^*_A,\label{eq:periodic-sifel}
\end{equation}
and subtracting corresponding terms on the opposite edges, we finally
obtain (compare with~\cite{Ozdemir:IJNME:2006})
\begin{eqnarray}
\left( \frac{\partial T}{\partial x} \right) L  & = & t_a - t_A = t_2 - t_1,\nonumber \\
\left( \frac{\partial T}{\partial y} \right) H  & = & t_b - t_B =
t_3 - t_1.\label{eq:constr-sifel}
\end{eqnarray}
These conditions can be introduced into most commercial software
products using the multi-point constraint equations.

  Note that neither the assumption of $t^*=0$ on $\Gamma$ nor the
  periodic conditions~\eqref{eq:periodic-sifel} (or
  constraints~\eqref{eq:constr-sifel}) are sufficient to specify the
  local temperature field uniquely. This condition is established by
  introducing a reference temperature $t(\tenss{x}^0)$ at an arbitrary
  point $\tenss{x}^0\in\Gamma$ as indicated in
  Eqs.~\eqref{eq:local_t-sifel}, recall also the opening
  Section~\ref{sec:fundamentals} together with
  Eqs.~\eqref{eq:bcmacro}$_1$ and~\eqref{eq:t}.  

In a linear case, when the effective thermal conductivities are
independent of the actual temperature, this value is arbitrary and can
be set equal to zero in the selected node of the finite element mesh.
In the temperature dependent problem the instantaneous effective
properties might, however, be strongly influenced by the current
temperature. This term then plays an important role.  In an example
presented in Section~\ref{subsec:comcodes} a reference temperature
equal to the selected initial temperature was assigned to one of the
corner nodes of the finite element mesh of the large RVE in
Fig.~\ref{fig:mesostrucutre}.  

\section{ Effect of moisture on thermal conductivity}\label{sec:moisture}
So far we restrained our attention to heat conduction problems under
steady state conditions. To that end, the balance equation, $q_{i,i} =
0$, Fourier's law and the equivalence of heat powers on micro- and
macro-scales, Eq.~\eqref{eq:tq}, were sufficient to derive both
effective conductivity and resistivity matrices.  When analyzing large
structural systems, such as historical stone bridges, cathedrals and
similar historical structures, the effect of moisture must be taken
into consideration and, therefore, the solution of the coupled heat
and moisture transfer is desirable.

\subsection{Computer codes used for coupled mass and energy transfer}\label{subsec:basicequations}
There exist a number of codes that allow for the description of
non-linear and non-stationary material behavior as well as cross -
effects when studying the moisture and heat transfer in heterogeneous
infrastructures. Here we recall at least two of them. The SIFEL
(SImple Finite Elements) computer code developed at our
department~\cite{Kruis:CC:2003} is a typical representative of
programs based on finite element techniques. This program was used to
derive the results presented in Section~\ref{subsec:comcodes}. The
DELPHIN program~\cite{Grunewald:1997,Delphin4} is on the other hand
developed on the bases of finite control volume method
(FCVM). Although not specifically used in this contribution this
program has proved useful in a number of verification tests.

In this section we mention two particular models implemented in the
SIFEL program. The model of Lewis and Schrefler is likely the most
advanced hygro-thermo-mechanical model so far employed for practical
calculations of transport processes in soils and other deformed porous
media. As its detailed description can be found
e.g. in~\cite{LewisSchrefler:98:2E}, we just summarize the basic
features of this model. The complete set of model equations comprises
the linear balance (equilibrium) equation formulated for a multiphase
body, the energy balance equation and the equations of mass conservation
for liquid water and gas. Assuming the rigidity of the solid phase,
the number of fundamental unknowns in the theory of the coupled
moisture and heat transfer reduces to three, namely the temperature
$t$ and the pressures in the liquid and gaseous phases $p^w$, $p^g$,
respectively. These unknown functions
can be solved from three basic equations
\begin{itemize}
\item one energy balance equation,
\item two equations of mass conservation.
\end{itemize}

The second model implemented in the SIFEL code is that of K\"{u}nzel
and Kiessl~\cite{Kunzel:IJHMT:1997}. It is much simpler then the
previous one and, despite the fact that the cross-effects are missing
in this model, it describes all substantial phenomena and its results
comply well with experimentally obtained data. For these reasons it
has been chosen to carry out the case study on masonry microstructure
and will be commented on in more detail in this paragraph. 

Vapor diffusion and liquid transport due to differences in capillary
suction stress are the main mechanisms of moisture transport. Although
an interaction of liquid and vapor fluxes under very humid
conditions cannot be excluded, they are treated as independent
processes in this model (vapor diffusion is most important in large
pores, whereas liquid transport takes place on pore surfaces, in
crevices and small capillaries).

Thermo-diffusion is neglected in this model and vapor flux, $\evek{{\it q}_v}$,
is described by the following equations
\begin{equation}
\evek{{\it q}_v} = -\delta_p\evek{\nabla{\it p}},\label{eq:dp}
\end{equation}
where $\delta_p$ is the vapor permeability of the porous material and
$p$ is the vapor pressure. The liquid transport incorporates the
liquid flow in the absorbed layer (surface diffusion) and in the water
filled capillaries (capillary transport). The driving potential in
both cases is capillary pressure (suction stress) and/or, according to
Kelvin's law, relative humidity $\varphi$. Hence, the flux of liquid
water, $\evek{{\it q}_w}$, can be written as
\begin{equation}
\evek{{\it q}_w} = -D_\varphi\evek{\nabla{\varphi}},\label{eq:Dj}
\end{equation}
where $D_\varphi=D_w\displaystyle{\frac{\de w}{\de\varphi}}$ is the
liquid conductivity 
, $D_w=D_w(w/w_f)$ is the
liquid diffusivity 
, $\de w/\de\varphi$ is the
derivative of the water retention function and $w/w_f$ is the water
content related to the capillary saturation. Finally, the heat flux,
$\evek{{\it q}_t}$, attains the following form
\begin{equation}
\evek{{\it q}_t} = -\chi\evek{\nabla{\it t}}-h_v\delta_p\evek{\nabla(\varphi{\it p}_{sat})},\label{eq:ht}
\end{equation}
where $h_v$ is the specific heat of evaporation and $p_{sat}$ is the
saturation vapor pressure.

In this model, the resulting set of differential equations for the
description of heat and moisture transfer, expressed in terms of
temperature and relative humidity $\varphi$, assumed the following
form:
\begin{itemize}
\item The energy balance equation
\begin{equation}
\left(\rho c + \frac{\de H_w}{\de t}\right)\dot{t}=\evek{\nabla}\trn\left[\chi
\evek{\nabla{\it
    t}}+h_v\delta_p\evek{\nabla(\varphi{\it
    p}_{sat})}\right]\label{eq:balance}
\end{equation}
\item The conservation of mass equation
\begin{equation}
\frac{\de
  w}{\de\varphi}\dot{\varphi}=\evek{\nabla}\trn\left[D_\varphi\evek{\nabla\varphi}+\delta_p\evek{\nabla(\varphi{\it
    p}_{sat})}\right].\label{eq:mass}
\end{equation}
\end{itemize}
In Eqs.~\eqref{eq:balance}~-~\eqref{eq:mass} $\rho$ is the material
density, $c$ is the specific heat capacity and $H_w$ is the enthalpy
of material moisture. Details concerning the specification of material
parameters can be found in~\cite{Kunzel:IJHMT:1997}. The storage
terms appear on the left hand side of Eqs.~\eqref{eq:balance}
and~\eqref{eq:mass}. The conductivity heat flux and the enthalpy flux
by vapor diffusion with phase changes in Eq.~\eqref{eq:balance} are
strongly influenced by the moisture fields. The vapor flux in
Eq.~\eqref{eq:mass} is governed by both the temperature and moisture
fields due to exponential changes of the saturation vapor pressure,
$p_{sat}$, with temperature. Vapor and liquid convection caused by
total pressure differences or gravity forces as well as the changes of
enthalpy by liquid flow are neglected in this model.

\begin{figure} [ht]
\begin{center}
\begin{tabular}{c}
\includegraphics[width=70mm,keepaspectratio]{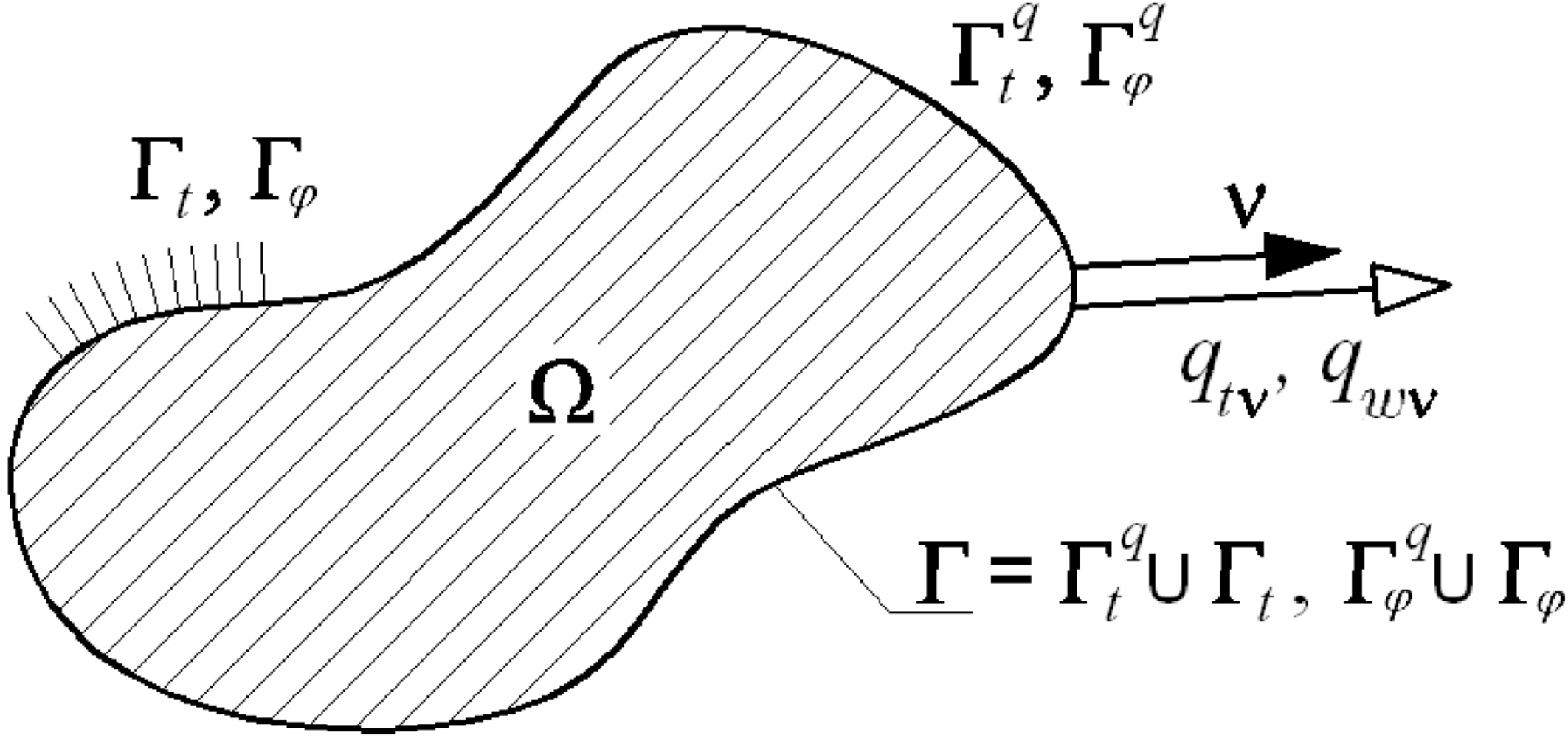}
\end{tabular}
\caption{A two dimensional region (e.g. an RVE in
  Fig.~\ref{fig:mesostrucutre} with prescribed heat flux)}
\label{fig:model}
\end{center}
\end{figure}

In the SIFEL program these equations are converted to a weak
formulation. 
Using the weighted residual statement the energy balance equation becomes
\begin{eqnarray}
&\int_\Omega w_t\left<\left(c+\frac{\de H_w}{\de t}\right)\dot{t}-
\evek{\nabla}\trn\left[\chi\evek{\nabla{\it t}}+h_v\delta_p\evek{\nabla(\varphi{\it p}_{sat})}\right]\right>\de\Omega+&\nonumber\\
&+\int_{\Gamma_t^q}w_t\left<\evek{\nu}\trn\left[\chi\evek{\nabla{\it t}}+
h_v\delta_p\evek{\nabla(\varphi{\it p}_{sat})}\right]+\overline{q}_{t\nu}\right>\de\Gamma = 0,\label{eq:weak-heat}
\end{eqnarray}
where $w_t$ is the weighting function such that $w_t=0$ on $\Gamma_t$,
see Fig.~\ref{fig:model}. Applying Green's theorem then yields
\begin{eqnarray}
&\int_\Omega \left\{w_t\left(c+\frac{\de H_w}{\de t}\right)\dot{t}+
\evek{\nabla{\it w}_t}\trn\left[\chi\evek{\nabla{\it t}}+h_v\delta_p\evek{\nabla(\varphi{\it p}_{sat})}\right]\right\}\de\Omega+&\nonumber\\
&+\int_{\Gamma_t^q}w_t\overline{q}_{t\nu}\de\Gamma = 0,\label{eq:weak-heat-2}
\end{eqnarray}
where $\overline{q}_{t\nu}$ is the prescribed heat flux perpendicular
to the boundary~$\Gamma_t^q$.  On $\Gamma_t$ the temperature $t$ is
equal to its prescribed value $\overline{t}$. Vector $\evek{\nu}$
stores the components of the unit outward normal defined already in
Section~\ref{sec:fundamentals}, see also Fig.~\ref{fig:model}.


The weak formulation for the conservation of mass can be derived analogously
\begin{eqnarray}
&\int_\Omega w_\varphi\left<\frac{\de w}{\de\varphi}\dot{\varphi}-
\evek{\nabla}\trn\left[D_\varphi\evek{\nabla\varphi}+
\delta_p\evek{\nabla(\varphi{\it p}_{sat})}\right]\right>\de\Omega+&\nonumber\\
&+\int_{\Gamma_\varphi^q}w_\varphi\left<\evek{\nu}\trn\left[D_\varphi\evek{\nabla\varphi}+\delta_p\evek{\nabla(\varphi{\it p}_{sat})}\right]
+\overline{q}_{w\nu}\right)\de\Gamma,
\label{eq:weak-moist}
\end{eqnarray}
where $w_\varphi$ is the corresponding weighting function such that
$w_\varphi=0$ on~$\Gamma_\varphi$. Application of Green's
theorem finally gives
\begin{eqnarray}
&\int_\Omega \left\{w_\varphi\frac{\de w}{\de\varphi}\dot{\varphi}+
\evek{\nabla{\it w}_\varphi}\trn\left[D_\varphi\evek{\nabla\varphi}+
\delta_p\evek{\nabla(\varphi{\it p}_{sat})}\right]\right\}\de\Omega+&\nonumber\\
&+\int_{\Gamma_\varphi^q}w_\varphi\overline{q}_{w\nu}\de\Gamma,
\label{eq:weak-moist-2}
\end{eqnarray}
where $\overline{q}_{w\nu}$ is the prescribed flux of
water perpendicular to the boundary~$\Gamma_\varphi^q$. On
$\Gamma_\varphi$ the relative humidity is again equal to its
prescribed value $\overline{\varphi}$.


A detailed specification of boundary conditions
used in our particular case will be given in the next section. For
details on numerical integration of Eqs.~\eqref{eq:weak-heat-2}
and~\eqref{eq:weak-moist-2} the interested reader is referred
to~\cite{Kruis:CC:2003}.

\subsection{Effective conductivities as a function of relative humidity and initial temperature}\label{subsec:comcodes}
\begin{figure} [ht]
\begin{center}
\begin{tabular}{cc}
\includegraphics[width=75mm,keepaspectratio]{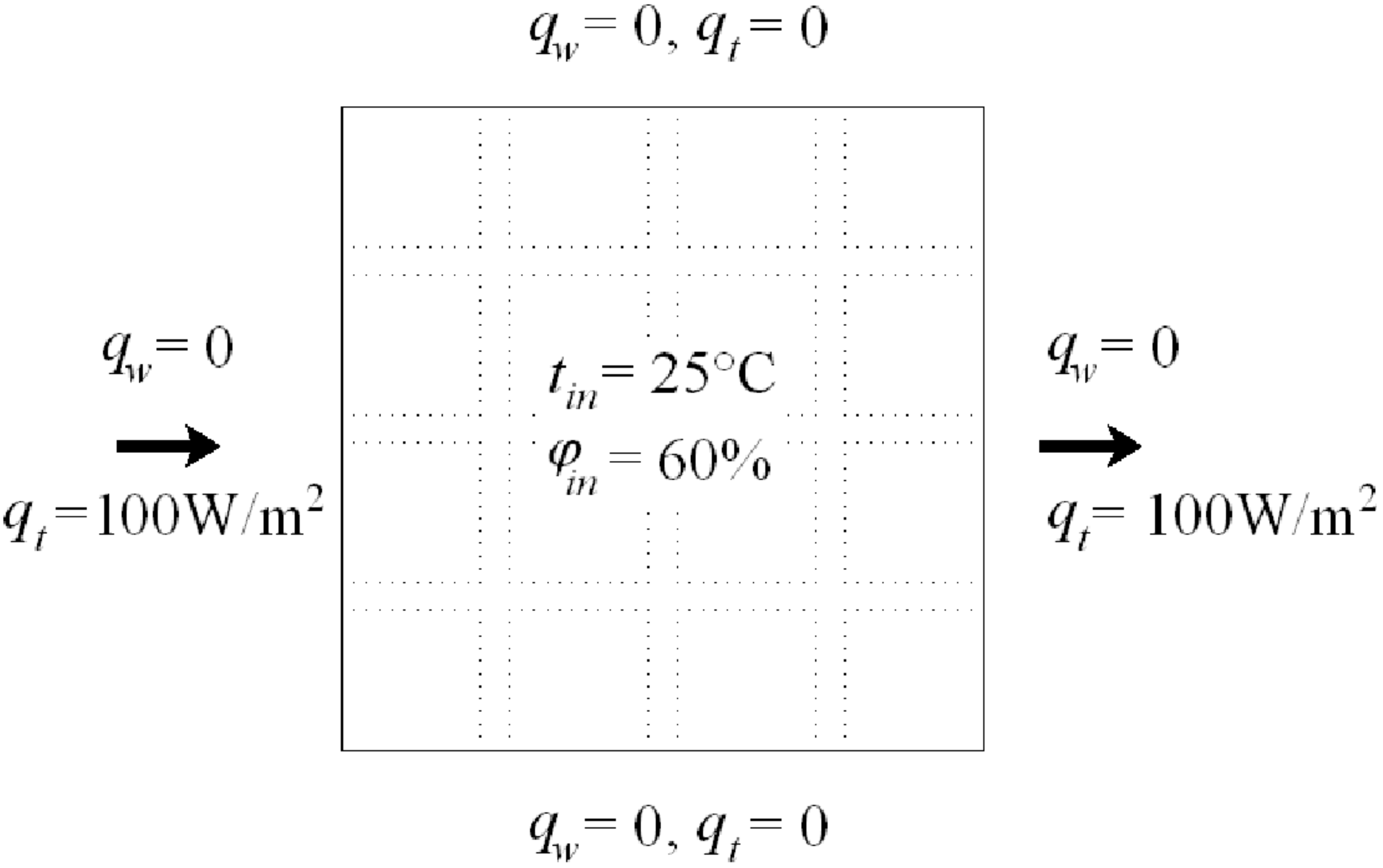}&
\includegraphics[width=65mm,keepaspectratio]{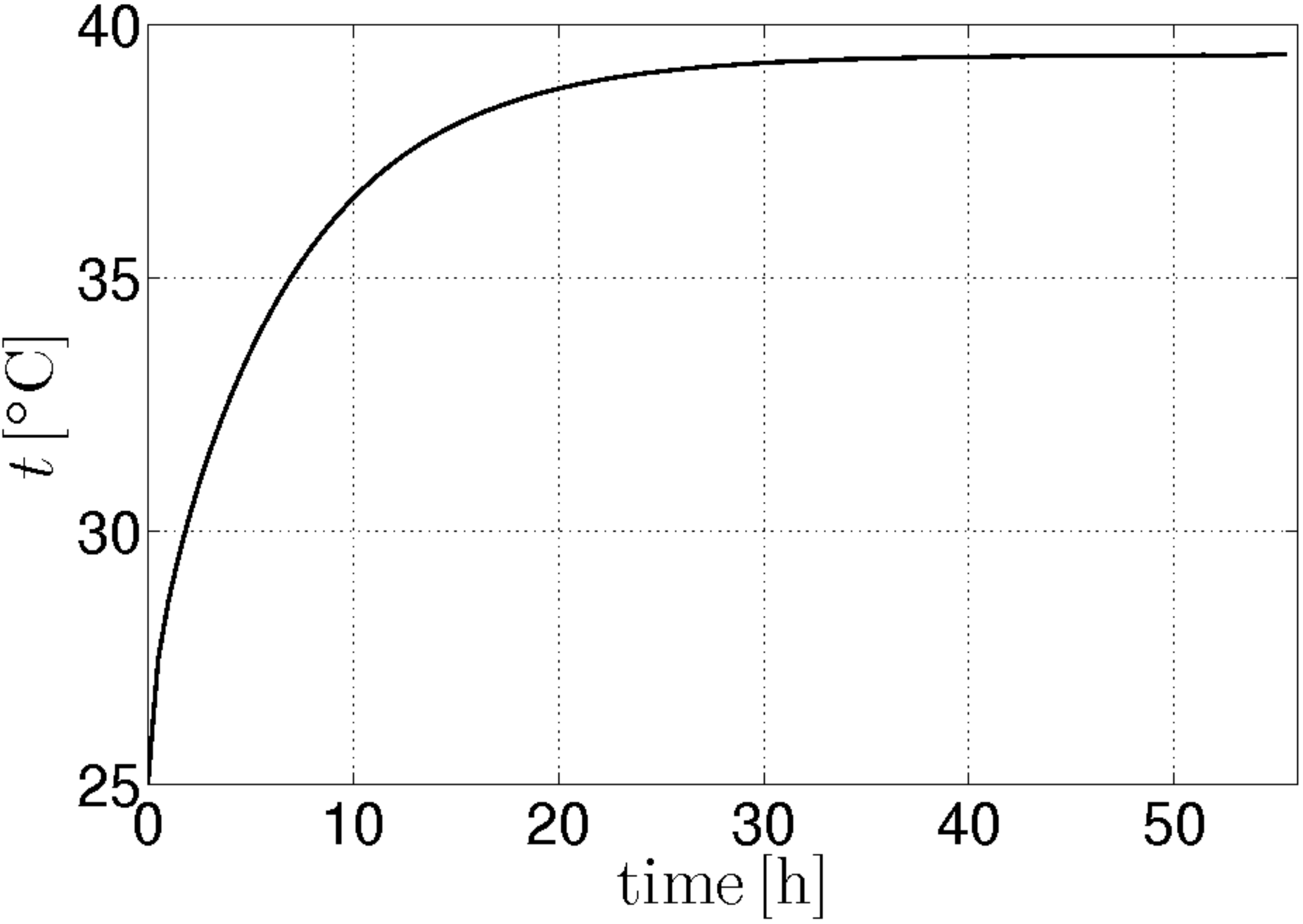}\\
(a)&(b)
\end{tabular}
\caption{Example: (a) Geometry and boundary conditions, (b) Evolution of temperature at $\tenss{x}^0$}
\label{fig:example}
\end{center}
\end{figure}

As an example the non-periodic 16-blocks RVE in
Fig.~\ref{fig:mesostrucutre} was examined to assess the influence of
moisture and initial temperature on the effective thermal
conductivities. Transient heat conduction problem was solved with the
help of SIFEL finite element program exploiting the K\"{u}nzel and
Kiessl model from the previous section. The loading and initial
conditions are plotted in Fig.~\ref{fig:example}(a). Although there are
no restrictions to introduce the periodic boundary conditions in the
SIFEL program through Eqs.~\eqref{eq:constr-sifel}, the present
analysis assumed, with reference to
Section~\ref{subsec:simpleexample}, the fluctuation part of
temperature $t^*=0$ on $\Gamma$. This simplification on the other hand
allows for a direct comparative study using other codes such as
DELPHIN, where implementation of periodic boundary conditions might be
precluded.

\begin{table}[!ht]
\begin{center}
\begin{tabular}{|l|c|c|}
\hline
Parameter & Mortar & Sandstone\\
\hline
density [kgm$^{-3}$] & 1700 & 1964\\
specific heat $c$ [Jkg$^{-1}$K$^{-1}$] & 1000 & 900\\
vapor diffusion resistance number $\mu$, see Eq.~\eqref{eq:mu} & 12 & 10\\
\hline
\end{tabular}
\end{center}
\caption{Constant material parameters of individual phases}
\label{Tab3}
\end{table}

\begin{figure} [ht]
\begin{center}
\begin{tabular}{cc}
\includegraphics[width=70mm,keepaspectratio]{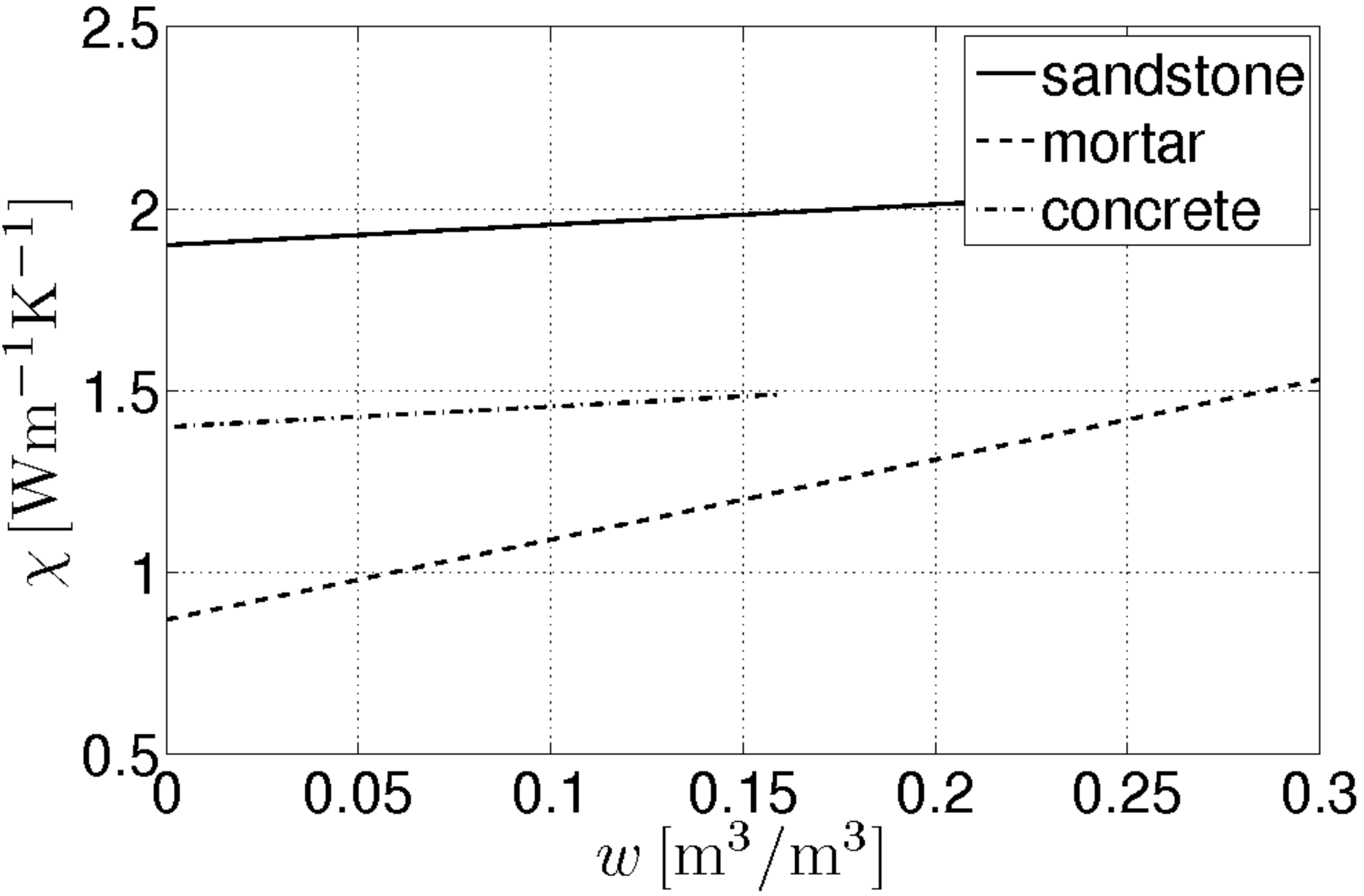}&
\includegraphics[width=70mm,keepaspectratio]{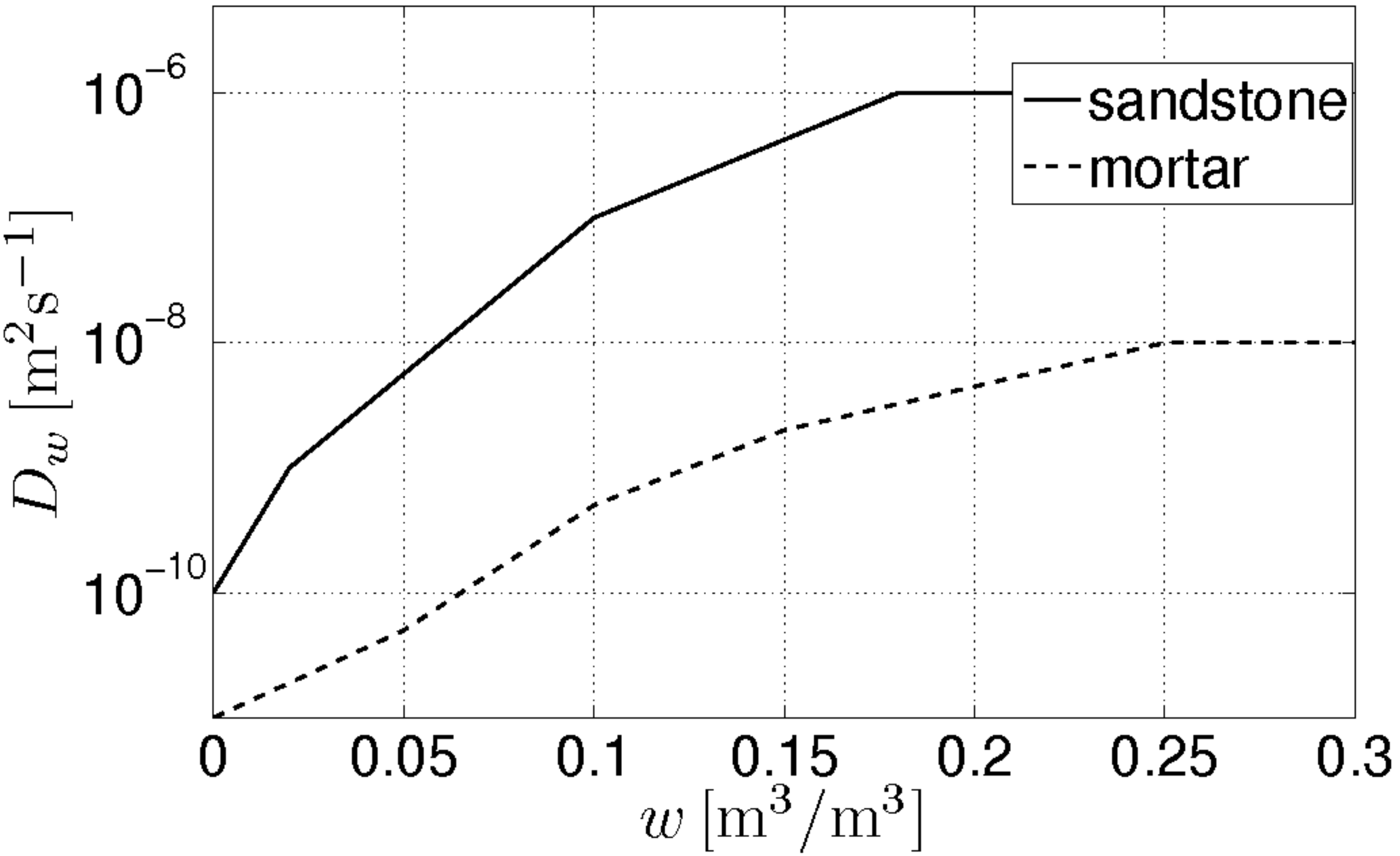}\\
(a)&(b)
\end{tabular}
\begin{tabular}{c}
\includegraphics[width=70mm,keepaspectratio]{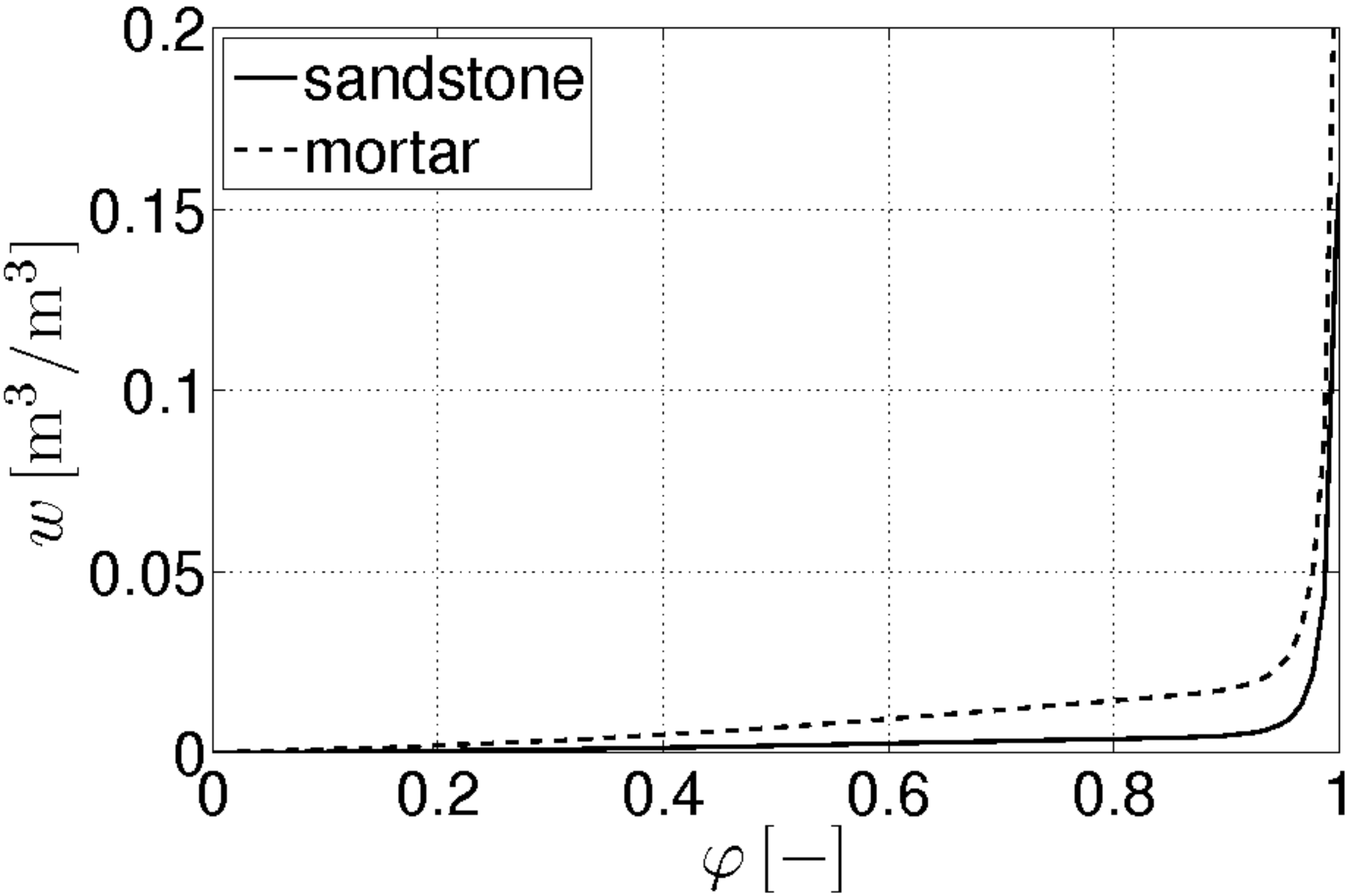}\\
(c)
\end{tabular}
\caption{Example - material data: (a) variation of phase thermal
  conductivities as a function of volume moisture, (b) variation of
  phase liquid diffusivities as a function of volume moisture, (c) phase
  sorption isotherms}
\label{fig:example-mater}
\end{center}
\end{figure}

The constant material parameters are listed in Table~\ref{Tab3}. The
moisture dependent variation of phase thermal conductivities and
liquid diffusivities is plotted in Fig.~\ref{fig:example-mater}
together with sorption isotherms obtained from the following
approximations~\cite{Delphin4}
\begin{eqnarray}
w&=&\left(1-\sqrt{1-\varphi}\right)\frac{w_{hyg}}{1-\sqrt{1-\varphi_{hyg}}},\\
w&=&w_{hyg}+\frac{\varphi-\varphi_{hyg}}{1-\varphi_{hyg}}\left(w_{sat}-w_{hyg}\right),
\end{eqnarray}
where the saturation moisture content $w_{sat}$ = 0.3, the hygroscopic
moisture content $w_{hyg}$ = 0.02 and the hygroscopic relative
humidity $\varphi_{hyg}$ = 95$\%$ were assumed.

\noindent
The remaining parameters that enter
Eqs.~\eqref{eq:dp}~-~\eqref{eq:mass} are provided by

\begin{itemize}
\item $h_v$ specific heat of evaporation [Jkg$^{-1}$]
\begin{eqnarray}
h_v = 2.5008 \cdot{} 10^6 \left ( \frac{273.15}{t}\right) ^ {(0.167 +
  t \cdot{} 3.67 \cdot{} 10^{-4})}.
\end{eqnarray}

\item $p_{\rm sat}$ saturation vapor pressure [Pa], see~\cite{Krejci:CTU:2001}
\begin{eqnarray}
p_{\rm sat} = {\rm exp} \left ( 23.5771 - \frac{4042.9}{t - 37.58}
\right ).
\end{eqnarray}

\item $\delta_p$ vapor permeability of the porous material
  [kgm$^{-1}$s$^{-1}$Pa$^{-1}$]
\begin{eqnarray}
\delta_p = \frac{\delta}{\mu},\label{eq:mu}
\end{eqnarray}
where $\mu$ is the vapor diffusion resistance number and $\delta$ is the vapor diffusion coefficient in air
[kgm$^{-1}$s$^{-1}$Pa$^{-1}$] given by, see~\cite{Schirmer:1938}
\begin{eqnarray}
\delta = \frac{2.306 \cdot{} 10^{-5 } \cdot p_a }{R_v \cdot t \cdot p}
\left ( \frac{t}{273.15} \right )^{1.81},
\end{eqnarray}
with $p$ set equal to atmospheric pressure $p_a$ = 101325 Pa and $R_v$
= $R$/$M_w$ = 461.5; $R$ is the gas constant (8314.41
Jkmol$^{-1}$K$^{-1}$) and $M_w$ is the molar mass of water (18.01528
kgmol$^{-1}$).
\end{itemize}

\begin{figure} [ht]
\begin{center}
\begin{tabular}{cc}
\includegraphics[width=70mm,keepaspectratio]{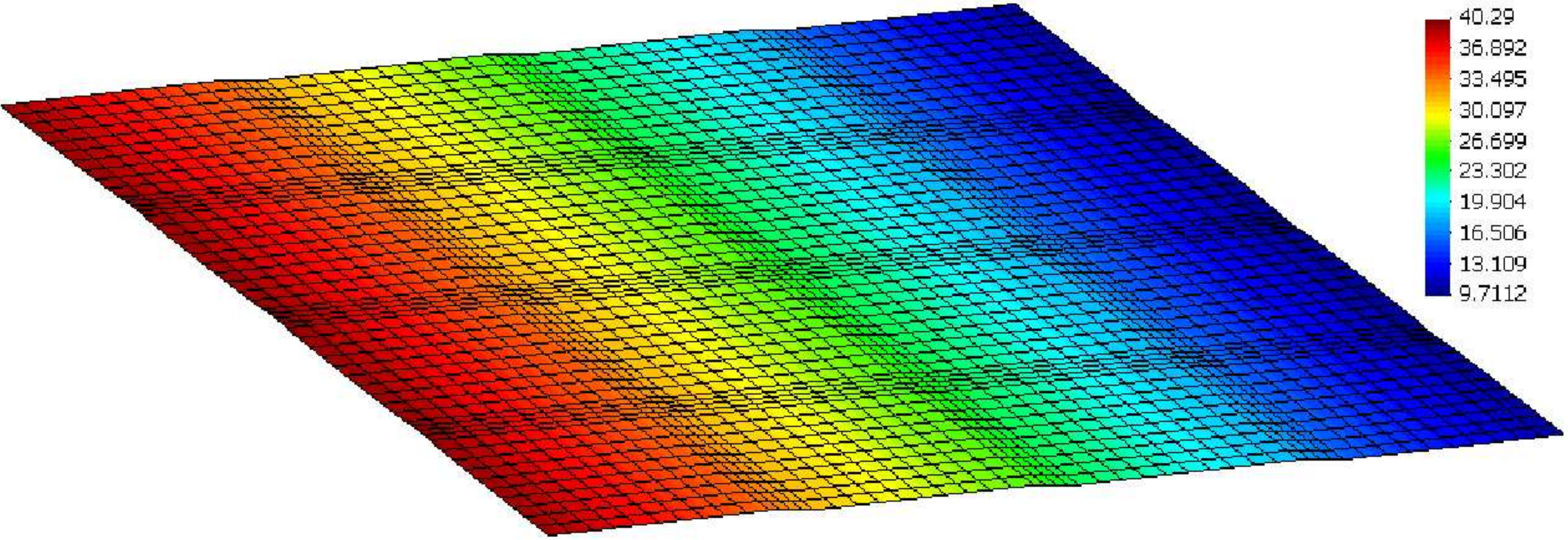}&
\includegraphics[width=70mm,keepaspectratio]{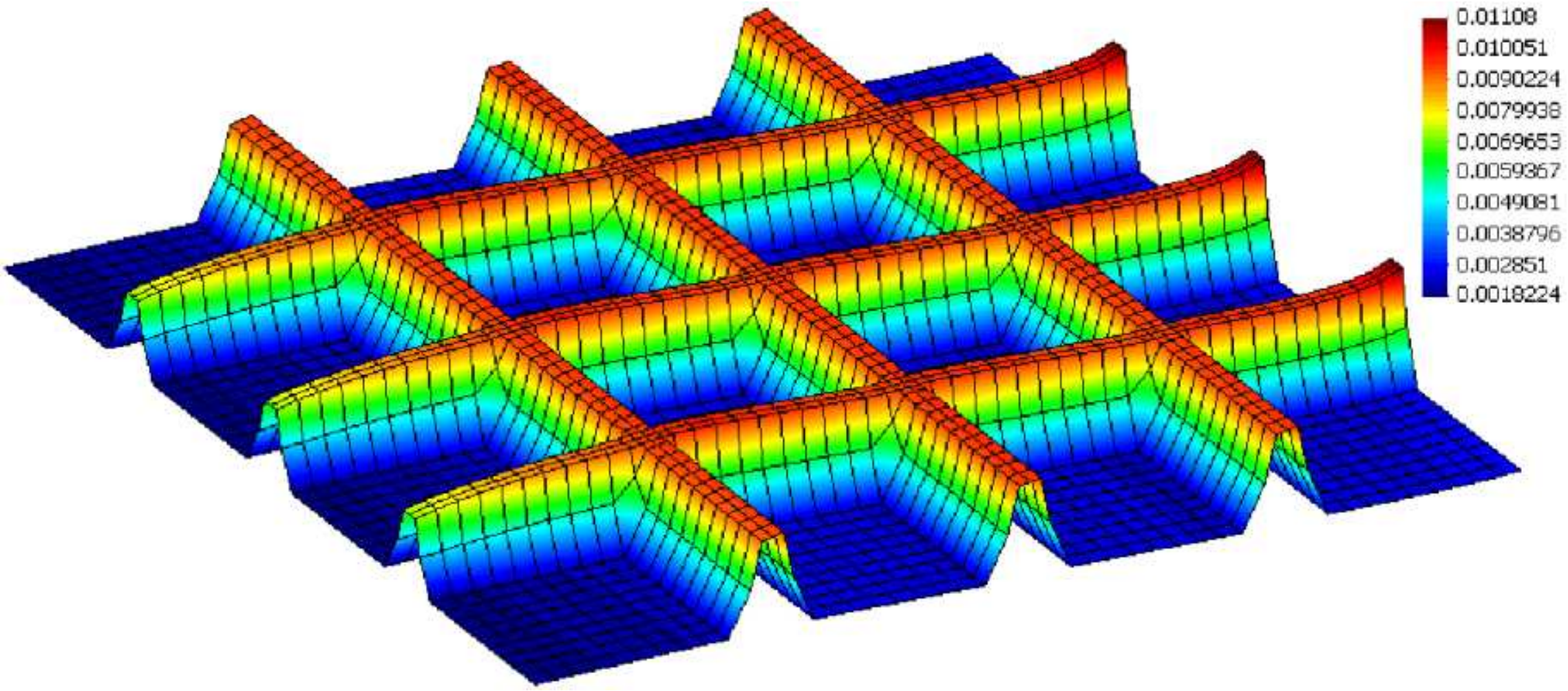}\\
(a)&(b)
\end{tabular}
\caption{Example: (a) Distribution of temperature (b) Distribution of volume moisture}
\label{fig:example-prubehy}
\end{center}
\end{figure}

The total temperature field $t = T + t^{*}$, which corresponds to
$q_{x}=\mathrm{100Wm^{-2}}$, appears in
Fig.~\ref{fig:example-prubehy}(a). In this case the room initial
temperature $t_{in}=25^{\circ}\mathrm{C}$ and relative humidity
$\varphi = 60\%$ were considered, recall Fig.~\ref{fig:example}(a).
As evident from Fig.~\ref{fig:example-prubehy}(b), the volume moisture
field $w\,\mathrm{(m^{3}/m^{3})}$ considerably varies throughout the
RVE thus affecting the macroscopic thermal conductivity slightly
depending on the size of the RVE.

\begin{figure} [ht]
\begin{center}
\begin{tabular}{cc}
\includegraphics[width=70mm,keepaspectratio]{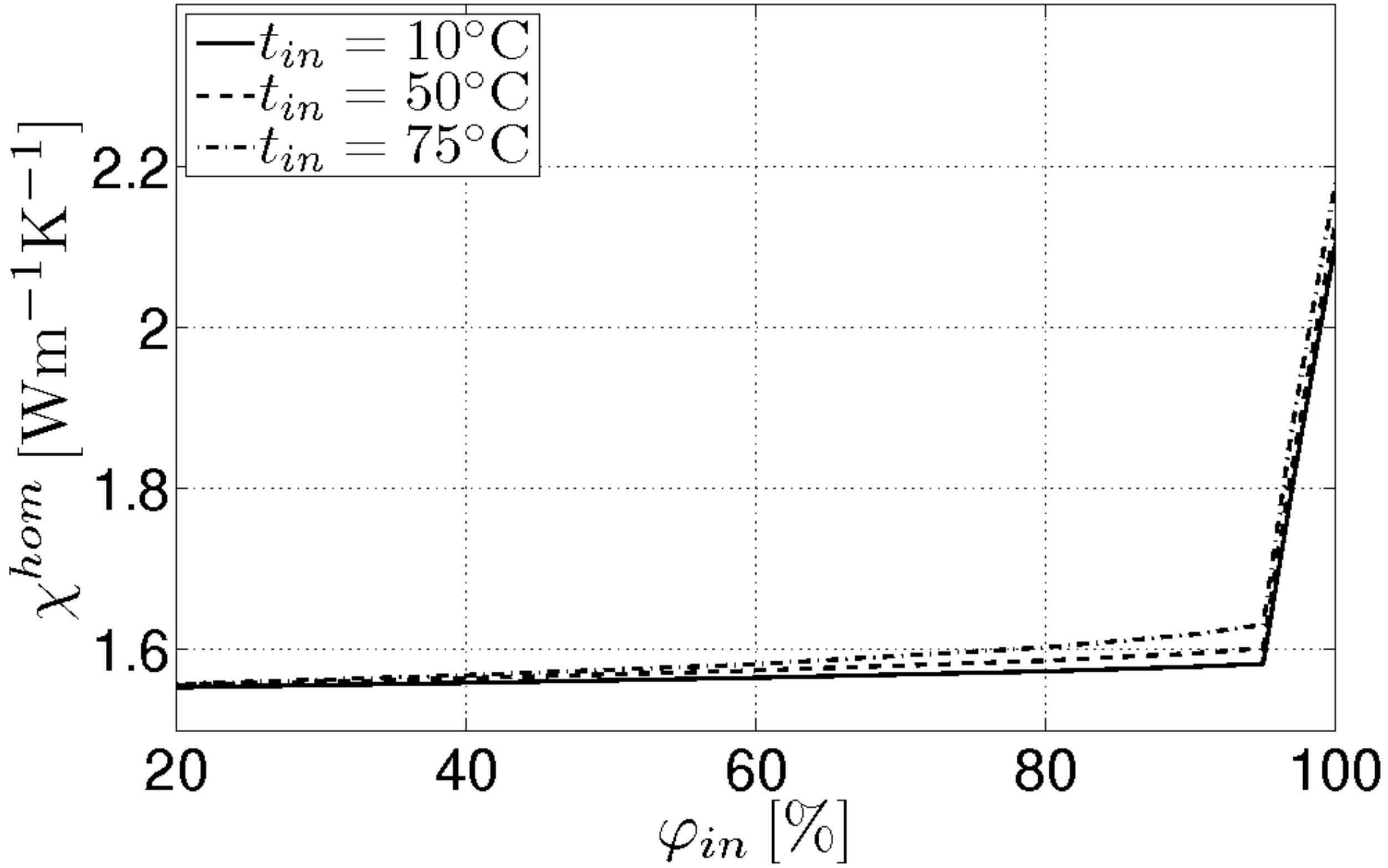}&
\includegraphics[width=70mm,keepaspectratio]{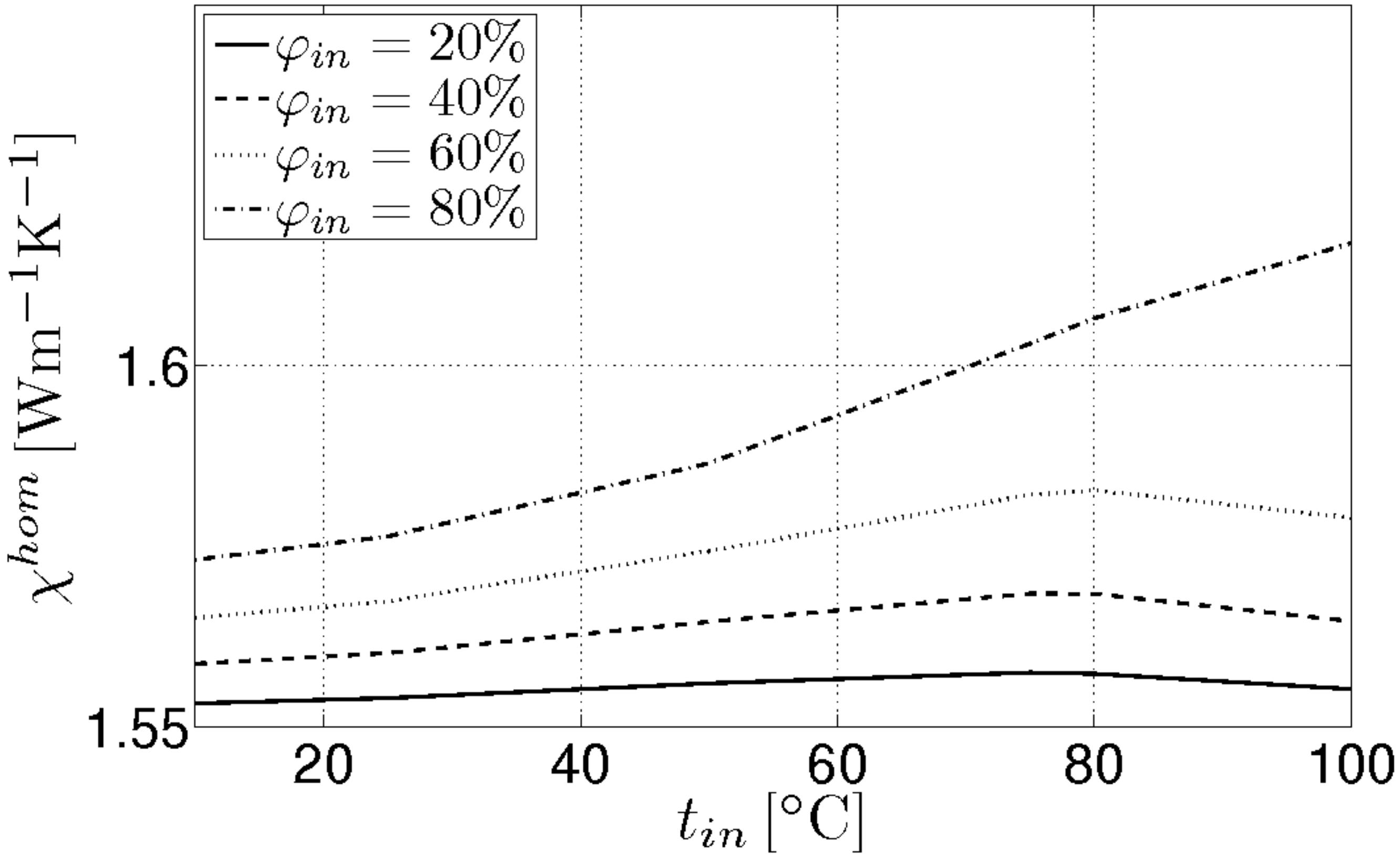}\\
(a)&(b)
\end{tabular}
\caption{Example - evolution of $\chi^{hom}$ as a function of: 
(a) $\varphi$ at $t_{in}=25^{\circ}\mathrm{C}$, 
(b) $t_{in}$}
\label{fig:example-grafy}
\end{center}
\end{figure}

Several interesting results have been derived within the scope of our
numerical experiments. The graph in Fig.~\ref{fig:example-grafy}(a)
displays the dependence of the macroscopic heat conductivity
$\chi^{hom}$ on the average relative humidity $\varphi$.  Individual
values of $\chi^{hom}$ follow again from
Eq.~\eqref{eq:chi-hom-gx}. Note that this equation corresponds to
steady state conditions attained after running the transient heat
conduction problem for about 50 hours, see Fig.~\ref{fig:example}(b)
showing an evolution of temperature at a reference point $\tenss{x}^0$
with respect to time.  It should be pointed out that the nearly linear
dependence between these two variables at a sub-hygroscopic region (up
to about 95\% - 98\%) becomes highly non-linear once the hygroscopic
moisture is attained and approaches the capillary and even maximum
(vacuum) saturation. Fig.~\ref{fig:example-grafy}(b) displays the
relation between the macroscopic thermal conductivity and the initial
temperature for different levels of the relative humidity
$\varphi_{in}$.

These information were subsequently utilized when performing the
analysis of Charles Bridge. It has been observed experimentally that
different sections of the bridge experience different moisture
content. The results in Fig.~\ref{fig:example-grafy} thus allowed us
to assign to each measured value $\varphi$ the corresponding value of
$\chi^{hom}$ in the macroscopic steady state analysis of Charles
Bridge. This then provided the space distribution of temperature
eventually used in the thermo-mechanical analysis of the
bridge~\cite{Novak:ES:2007}.

\begin{figure} [ht]
\begin{center}
\includegraphics[width=70mm,keepaspectratio]{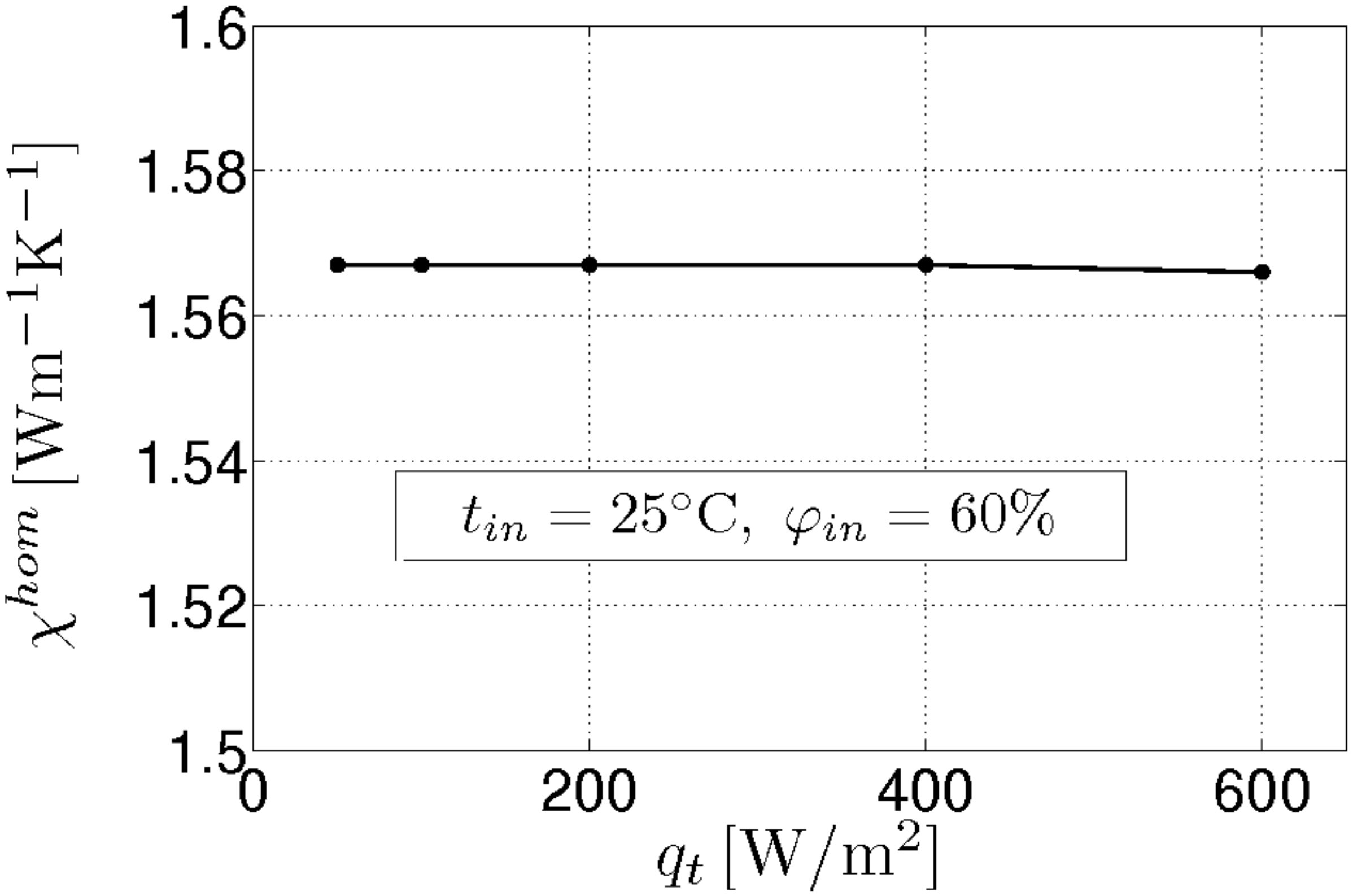}
\caption{Example: sensitivity of $\chi^{hom}$ on variation of $T_{,i}$}
\label{fig:example-sens}
\end{center}
\end{figure}

The sensitivity of effective thermal conductivities on the variation
of macroscopic temperature gradient was also studied. Clearly, the
results presented in Fig.~\ref{fig:example-sens} suggest that the
coefficient of the macroscopic thermal conductivity $\chi^{hom}$ is
almost insensitive to the changes in the macroscopic temperature
gradient $T_{,i}$, which considerably simplifies the micro
(meso)-macro approach.

\section{Conclusions}\label{sec:conclusions}
Homogenization techniques applied to a meso-scale provide the values
of effective transport parameters, such as the coefficient of thermal
conductivity (and/or moisture permeability), and allow for their
dependence on both macroscopic temperature and moisture fields. As
efficient tools, the SIFEL and DELPHIN computer codes can be
effectively used when analyzing transport processes within a certain
PUC to get distributions of these fields on a meso-scale, while taking
into account the coupling effects in transport phenomena. With reference
to thermal conductivity the following conclusions can be pointed out:
\begin{itemize}
\item At low levels of moisture the coefficient of thermal
conductivity varies almost linearly with the relative humidity
$\varphi$. This relation becomes strongly non-linear as soon as the
hygroscopic moisture $(\varphi\approx 95\%)$ is attained.
\item The macroscopic volume moisture
$(\mathrm{m}^{3}/\mathrm{m}^{3})$ extensively varies throughout
the RVE thus affecting the value of macroscopic thermal
conductivity.
\item There exists a certain dependence of the macroscopic thermal
conductivity on the initial temperature $t_{in}$. But it is
distinctive just at higher temperatures, say above
$30^{\circ}\mathrm{C}$.
\item A very important finding is that the macroscopic
conductivity is nearly insensitive to the macroscopic temperature
gradient, even in the non-linear range, which simplifies the
meso-macro approach.
\end{itemize}

\begin{figure}
\centering \SEPUCfig{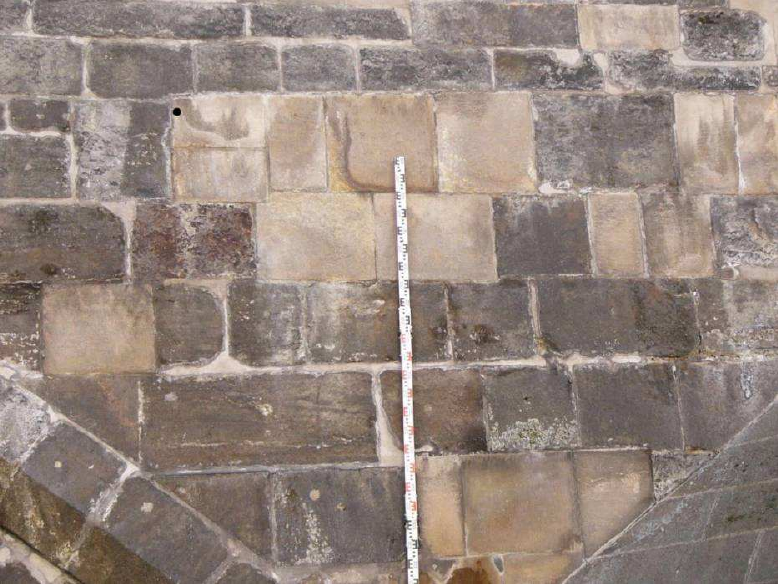}{.32}{.32}{.225}
\caption{Construction of statistically equivalent periodic unit cell
  (SEPUC) for random-coursed masonry~\cite{Zeman:MSMSE:2007}}
\label{F:masonry}
\end{figure}

Finally recall a relatively good correspondence between the
numerically obtained values of the effective conductivities and those calculated
from simple rules of mixture, see Table~\ref{Tab2}. It should be,
however, mentioned that such a good agreement is merely attributed to
the selected geometrical arrangement of the two phases. When more
realistic periodic unit cells (such as the one displayed in
Fig.~\ref{F:masonry}) were used, the simple bounds were too far
apart to be of any practical use.

\section*{Acknowledgment}
This outcome has been achieved with the financial support of the
Ministry of Education, Youth and Sports, project No. 1M0579, within
activities of the CIDEAS research centre. In this undertaking,
theoretical results gained in the project GACR 103/04/1321 and
103/08/1531 were partially exploited.


\end{document}